\documentclass[12pt]{article}

\usepackage{latexsym}
\usepackage{amssymb,amsfonts,amsmath}
\usepackage{graphicx}
\usepackage{indentfirst}
\usepackage{color}
\usepackage{amsfonts}

\parskip=\medskipamount

\newcommand {\cD}{{\cal D}}
\newcommand {\cE}{{\cal E}}

\newcommand {\cN}{{\cal N}}

\newcommand {\cR}{{\cal R}}

\def\a{\alpha}

\def\b{\beta}
\def\c{\chi}
\def\d{\delta}
\def\e{\epsilon}
\def\f{\phi}
\def\g{\gamma}
\def\G{\Gamma}

\def\l{\lambda}

\def\o{\omega}

\def\q{\theta}
\def\r{\rho}
\def\s{\sigma}

\def\z{\zeta}
\def\D{\Delta}
\def\F{\Phi}
\def\J{\Psi}

\def\O{\Omega}

\def\S{\Sigma}
\def\U{\Upsilon}
\def\X{\Xi}

\def\rd{{\rm d}}
\def\ri{{\rm i}}
\def\re{{\rm e}}

\def\gg{{\hat g}}

\newcommand{\ad}{{\dot{\alpha}}}                           
\newcommand{\bd}{{\dot{\beta}}}                            
\newcommand{\ve}{\varepsilon}                            
\newcommand{\cDB}{{\bar\cD}}                            

\newcommand{\ab}{{\a\b}}

\newcommand{\pa}{\partial}                           
\newcommand{\hf}{\frac12}

\newcommand{\vf}{\varphi}

\newcommand{\be}{\begin{equation}}
\newcommand{\ee}{\end{equation}}
\newcommand{\bea}{\begin{eqnarray}}
\newcommand{\eea}{\end{eqnarray}}
\newcommand{\non}{\nonumber}
\def\double #1{#1{\hbox{\kern-2pt $#1$}}}

\newcommand{\gd}{{\dot\g}}

\newif\ifdtup

\newcommand{\bsubeq}{\begin{subequations}}
\newcommand{\esubeq}{\end{subequations}}

\numberwithin{equation}{section}

\begin{document}

\begin{titlepage}
\begin{flushright}
September, 2019 \\
\end{flushright}
\vspace{5mm}

\begin{center}
{\Large \bf Comments on Anomalies in Supersymmetric Theories}
\\

\vspace{25pt}
{\large Sergei M. Kuzenko$^{a}$,  Adam Schwimmer$^{b}$ and Stefan~Theisen$^{c}$}

\vspace{10pt}
{\sl\small
$^a${\it Department of Physics M013, The University of Western Australia\\
35 Stirling Highway, Perth W.A. 6009, Australia}
}

\vspace{10pt}
{\sl\small
$^b${\it  Weizmann Institute of Science, Rehovot 76100, Israel}
}

\vspace{10pt}
{\sl\small
$^c${\it Max-Planck-Insitut f\"ur Gravitationsphysik, Albert-Einstein-Institut, 14476 Golm, Germany}
}

\end{center}

\begin{abstract}
\baselineskip=14pt

We analyse the relation between anomalies in their manifestly supersymmetric formulation in superspace and
their formulation in Wess-Zumino (WZ) gauges.
We show that there is a one-to-one correspondence between the solutions of the cohomology problem in the two
formulations and that they are related by a particular choice of a superspace counterterm (``scheme'').
Any apparent violation of $Q$-supersymmetry is due to an explicit
violation by the counterterm which defines the scheme equivalent to the
WZ gauge. It is therefore removable.
\end{abstract}
\vspace{5mm}

\vfill

\vfill
\end{titlepage}

\newpage
\renewcommand{\thefootnote}{\arabic{footnote}}
\setcounter{footnote}{0}

\tableofcontents{}
\vspace{1cm}
\bigskip\hrule

\allowdisplaybreaks

\section{Introduction}
Anomalies in supersymmetric theories were understood in the superspace formulation a long time ago.
Having a supersymmetric theory one can couple it to external source  superfields. The Ward identities
related to global symmetries of the microscopic theory can be equivalently studied by analyzing the gauge invariances of the
effective action, which  depends on the sources after the microscopic fields are integrated out.
The two typical examples which we will discuss in this paper are microscopic theories  with ``flavour" symmetries
and superconformal symmetries. In these cases the sources are gauge super-fields and the  Einstein supergravity multiplet,
respectively.
The gauge groups in superspace are (super)Lie groups. For the flavour symmetry each Lie group generator is parameterized
by a chiral superfield while for the superconformal theories we have a semidirect product of superspace
reparametrizations with super-Weyl transformations. The former are parameterized by vector and spinor holomorphic superfields
and the latter  by a holomorphic scalar superfield.

The local anomalies we are discussing are related to operators in the microscopic theory which vanish on-shell.
For lack of a better name we will call them null-operators:
divergence of a current, trace of the energy-momentum tensor, fermionic and auxiliary components in the
respective anomaly multiplets, etc.
One would naively expect
that all correlators in the microscopic theory which involve the null-operators vanish, i.e. that these operators decouple
from the theory.
Since the operators are on-shell, this is automatic for the imaginary parts of the correlators.
However in certain correlators the real parts,
which are necessarily present by analyticity, cannot be chosen to satisfy the Ward
identities which follow from decoupling. This is the anomaly.
Since real parts can be added arbitrarily, anomalous correlators are always defined
modulo arbitrary polynomials in the external momenta, whose choice defines  a ``scheme".
Changing the scheme may change the overall symmetry preserved by the correlators, but there is no choice where
all Ward identities are non-anomalous.

The above statements have a clear translation into the generating functional  formalism, where ``real parts" correspond
to local terms in the external gauge fields.
One can add local terms to the generating functional defining ``the scheme", while the non-local piece
corresponds to the imaginary part.
The anomaly corresponds to a local gauge variation of the generating functional which cannot be eliminated by a
choice of scheme, i.e. by adding a local term to the generating functional. This defines a cohomology problem,
whose nontrivial solutions are the anomalies. By adding suitable counterterms the anomaly can be shifted between
different symmetries, but it cannot be eliminated altogether.

Anomalies of supersymmetric theories were completely analyzed and explicit local
superspace expressions were given, as discussed in detail in  \cite{PSS,PS,BPT} and more recently in \cite{GG}.
Furthermore, the impossibility of removing them by adding local counterterms was proven.
The superspace expressions imply a particular scheme which respects, by construction,
supersymmetry and additional subgroups of the gauge symmetry:
transformations with constant gauge parameters
for the ``flavour" symmetry and superspace reparametrizations for the superconformal case.

For the supersymmetric anomalies there arises a special situation:
the gauge symmetries can be partially fixed in an ``ultralocal" fashion:
the gauge fixing is done on the $\theta$ dependence, but it is completely algebraic in
$x$-space, i.e.  it does not involve derivatives in $x$ space.\footnote{A similar situation exists in
nonsupersymmetric theories for the trace anomalies: one can choose a gauge where the trace of the energy-momentum tensor is
identically zero and one deals just with the Ward identities following from conservation. This will be
further discussed in Appendix B.}
As a consequence one can study the anomaly problem
in a meaningful fashion in these gauges, called generically ``WZ gauges" in the following.
The exact relation between the anomalies in these WZ gauges, including the relation to the original
manifestly supersymmetric expressions for the anomaly in
superspace, is the topic of our discussion.

For notational convenience,  we start with a discussion of  the set-up in a general framework.
Consider the generating functional $\Gamma(A)$ where
$A$ is acted upon by elements $G$ of the full gauge group $\mathcal{G}$.
The $x$ and $\theta$ dependence of the fields $A$ and group elements $G$ is left out,  again for the sake of notational simplicity.
In our case $A$  represents the full set of  superspace gauge fields and
$\mathcal{G}$ the full gauge group in superspace. Consider now a partial  gauge fixing
to configurations $\bar{A}$ and denote the residual gauge group by $\bar{\mathcal{G}}$.
Each element $G \in \mathcal{G}$ can then be represented in terms of an element of the coset
$\mathcal{G}/\bar{\mathcal{G}}$ times an element of $ \bar{\mathcal{G}}$.
This decomposition is generically ambiguous
since we can multiply the element of the coset with an arbitrary element of $\bar{\mathcal{G}}$ and the
group element of $\bar{\mathcal{G}}$ with its inverse.

In the supersymmetric context we are in a special situation: in WZ gauges the gauge fixing occurs by
ultralocal  (super)gauge transformations which do not involve derivatives with respect to $x$,
i.e. the coset element $G_0(A)$ is local.
Its defining property is the relation between the original configurations $A$ and
the gauge fixed ones $\bar{A}$:
\begin{equation}\label{local}
\bar{A}= A^{G_0(A)}\,.
\end{equation}
The condition of locality of $G_0(A)$ and the fact that
the relation \eqref{local} between the configurations $A$ and $G_0(A)$ is one to
one due to the ultralocality, define the situation for which  our general discussion below applies.
For configurations in the WZ gauge $G_0(\bar A)=e$, where $e$ is the unit element of the gauge group.

In terms of the microscopic theory the above gauge choice amounts to putting to zero correlators involving
``ultralocal null operators". They couple to sources whose gauge
transformation is algebraic. They can be removed by a choice of WZ gauge.
Different  WZ gauges are characterized by the set of ultralocal null operators which were put to zero.
In contrast to this, one keeps
all ``divergence null operators",
i.e. divergences of currents which couple to sources with a differential gauge transformation.
Their sources survive in WZ gauge.

The group manipulations are valid in the specific representation where the group acts on $A$: as a consequence
we will have group parameters depending on $A$. Here it is essential that
this dependence is local in $x$ space and therefore the  anomaly analysis makes sense also for the gauge fixed situation.

We start by discussing the relation between the generating functional in superspace $\Gamma(A)$ and
in the fixed gauge $\bar{\Gamma}(\bar{A})$ in the simplest situation, i.e. when there are no anomalies.
Then the relation is trivial. Starting with $\Gamma$ we get:
\begin{equation}\label{direct}
\bar{\Gamma}(\bar{A})=\Gamma(A=\bar{A})
\end{equation}
Conversely, if $\bar{\Gamma}(\bar{A})$, the generating functional in the Wess-Zumino gauge
is known, we can reconstruct the full $\Gamma(A)$ by simply defining:
\begin{equation}\label{inverse}
\Gamma(A)=\bar{\Gamma}(A^{G_0(A)})
\end{equation}
The above relations  express the fact that in the situation of non-anomalous gauge invariance the generating
functional really depends on the gauge orbit  and $\bar{A}$ is an unambiguous label of the orbit.

We now discuss the situation when anomalies are present.
Then for a generic  supergauge transformation $G$
the generating functional $\Gamma$ is no longer invariant but obeys the anomalous transformation rule
\begin{equation}\label{WZgen}
\Gamma(A^G)=\Gamma(A) + W(A;G)\,,
\end{equation}
where $W(A;G)$ is a local functional of $A$ and $G$, called the ``Wess-Zumino functional" in the following.
It represents a solution to the WZ cohomology problem.
It cannot be written as a $G$ transformation  of a local functional  of $A$. For infinitesimal $G$  \eqref{WZgen} gives
the anomaly, i.e. \eqref{WZgen} represents the integrated form of the anomaly.
Since the l.h.s. of \eqref{WZgen} gives a representation of the gauge group $\mathcal{G}$,  $W(A;G)$ obeys
the consistency condition (``WZ condition"):
\begin{equation}\label{WZcond}
W(A;G_1 G_2)=W(A;G_2) +W(A^{G_2};G_1)\,.
\end{equation}
Given the anomaly, $W$ always exists but its explicit form is not always available.
For abelian flavour symmetries
and the superconformal case explicit expressions can be written down.

Since the anomalous transformation involves the local quantity $W$, the imaginary parts are not affected by it.
Therefore the imaginary parts of the generating functionals $\Gamma$ and $\bar{\Gamma}$ continue to be
related by the naive relations \eqref{direct} and \eqref{inverse}.
Once the full generating functionals are considered,
the relation is no longer so simple. In particular
in the WZ gauge we have a new cohomology problem for the residual gauge transformations,
i.e. we have functionals  $\bar{\Gamma}(\bar{A})$
which should satisfy the conditions \eqref{WZgen} and \eqref{WZcond} with  $A$ and $G$
replaced by $\bar{A}$ and $\bar{G}$, respectively, for a local functional $\overline{W}(\bar{A};\bar{G})$.
Moreover the gauge transformations relating a general configuration to its WZ-representative could be anomalous
such that \eqref{direct} and \eqref{inverse} are not applicable.

In one direction the relation is  simple: given a $\Gamma(A)$ which obeys the condition \eqref{WZgen},
the restriction to the WZ gauge
\begin{equation}\label{rest}
\bar{\Gamma}_r(\bar{A})\equiv \Gamma(A=\bar{A})
\end{equation}
will give a solution of the WZ cohomology problem  with
\begin{equation}\label{restW}
\overline{W}_r(\bar{A};\bar{G})\equiv W(A=\bar{A};G=\bar{G})
\end{equation}
In the opposite direction the relation is equally straightforward: assume one has a solution of the cohomology problem
for a $\bar{\Gamma}(\bar{A})$ with a corresponding
$\overline{W}(\bar{A};\bar{G})$. Then we can   define a generating functional for arbitrary configurations in superspace by:
\begin{equation}\label{up}
\Gamma_{u}(A)\equiv \bar{\Gamma}(A^{G_{0}(A)})\,.
\end{equation}
Since  the group element $G_{0}(A^G ) G {[G_{0}(A)]}^{-1}$ belongs to $\bar{\mathcal{G}}$, it is easy to show that
$\Gamma_u$ defined by \eqref{up} obeys \eqref{WZgen} with $W_u$ given by
\begin{equation}\label{upW}
W_u (A;G)\equiv\overline{W}( A^{G_0(A)};G_{0}(A^G ) G {[G_{0}(A)]}^{-1})\,.
\end{equation}
By construction $\Gamma_u$, though formally defined  on the full superspace configuration $A$,
after the superspace integration depends only on the components of $A$ present in the Wess-Zumino
gauge. It represents the mapping of the cohomology in the Wess-Zumino gauge to the full superspace.
We assumed that in superspace all the solutions of the cohomology are known and are represented by $\Gamma(A)$.
Therefore  $\Gamma_u$ defined above should differ from it by a local functional of $A$.

After establishing the isomorphism  of the cohomologies in superspace and in WZ gauge we would like to be  
more specific and relate directly the generating functionals. This is possible only if the WZ gauge represents a genuine gauge 
fixing, i.e. the gauge direction represented by the choice $G_0(A)$ is anomaly free. We can achieve this by choosing a 
new scheme. Define a new generating functional $\tilde{\Gamma}(A)$
\begin{equation}\label{uppp}
\tilde\Gamma(A)=\Gamma(A)+W(A;G_{0}(A))=\Gamma(\bar A)\,.
\end{equation}
The new generating functional obeys by construction the identity
\begin{equation}\label{indep}
\tilde{\Gamma}(A^{G_{0}(A)})=\tilde{\Gamma}(A)\,,
\end{equation}
i.e. now \eqref{direct} becomes  a genuine gauge fixing.
While $\Gamma(A)$ and $W(A;G_{0}(A))$ each depend on all the components of $A$, their combination appearing in
\eqref{uppp} depends just on the components of the
WZ gauge and the dependence on them coincides
with $\Gamma(\bar{A})$.

We have therefore a general procedure to map the generating functional in a WZ gauge to a particular scheme
in superspace.  This was achieved by adding the local counterterm  $W(A;G_{0}(A))$. This counterterm
may violate additional symmetries being at the origin of the apparent violations in the WZ gauge.
The above pattern, i.e. transforming the WZ gauge into a particular scheme in superspace, will  be used in all the examples
discussed in the paper.

This general discussion led us to the conclusion that due to the special properties of the WZ gauges, the
cohomology problems in full superspace and in the WZ gauges are completely equivalent.
In particular to any anomaly solution in the WZ gauge corresponds in superspace a local counterterm,
i.e.  a particular  allowed scheme. Therefore if in a WZ gauge a particular symmetry is violated compared
to the superspace formulation, it just means that a particular counterterm violating the symmetry was added and
removing it will lead to a symmetric formulation making the apparent violation spurious.

In all the cases discussed in this paper the cohomology in the WZ gauge is given by $\bar\Gamma_r$
defined  in \eqref{rest}.
Therefore using \eqref{uppp} the counterterm in superspace is simply $W(A;G_{0}(A))$  giving  a very simple
realization of the ``anomaly shifting" paradigm. The symmetries and the anomalous Ward identities
in the new scheme can be directly  obtained from the properties of the counterterm  $W(A;G_{0}(A))$ .

As a general conclusion the  anomalies seen in the WZ gauge cannot have an absolute meaning.
In particular any additional anomaly compared to the standard superspace anomalies can be removed by
simply removing $W$, i.e. adding the local counterterm  $-W(A;G_{0}(A))$.

In the paper we will analyze in detail three examples which fit into the general pattern described above:
we will identify in each example the fixed gauge space and the residual gauge group.
We will specify the WZ functionals giving the local counterterms  for each case and analyze the symmetries apparently broken.

The rest of the paper is organized as follows:
in Section 2 we analyze the anomalies in  supersymmetric models  with a global $U(1)$
symmetry, tracing the apparent violation of global SUSY in the WZ gauge.
In Section 3 we review the formulation of super-Weyl anomalies in superspace.
In Section 4 we discuss a gauge (equivalently a scheme) which is minimal for having a non-anomalous $Q$-supersymmetry.
In Section 5 the WZ gauge and its associated scheme, which have an apparent anomaly in $Q$-supersymmetry, is analysed.
We identify the direction which became non-anomalous when the anomaly in $Q$-supersymmetry appeared.
In the concluding Section 6 we summarize  the relations between WZ gauges and respective schemes and we discuss the
general pattern of the interplay of $Q$-supersymmetry with other symmetries.

In two Appendices we discuss the construction of WZ actions and WZ-like gauges in
nonsupersymmetric theories, respectively.

In this paper we discuss ${\cal N}=1$ supersymmetric theories in four dimensions, using heavily their superspace
formulation. General references are \cite{WB},\, \cite{GGRS} and \cite{BK}. We will largely follow the
notation and conventions of the first and third of these references.


\section{Flavour anomaly}

In this section we mostly review well known facts. They are discussed in detail in \cite{Nair,Guadagnini} and
more recently in \cite{Closset}.
Consider a supersymmetric field theory with global (``flavour") symmetries.
For simplicity we only discuss the abelian case when the symmetry is $U(1)$.
In the references the non-abelian case is also considered.

There is an associated Noether current $J=J^\dagger$ which is classically conserved on-shell:
\begin{equation}\label{conservation}
D^2 J=\bar D^2 J=0~.
\end{equation}
For instance, for the free massless WZ model $J=\Phi^\dagger\Phi$. The conservation equation means that $J$ is a linear multiplet on-shell, 
i.e. its higher components ($\theta^2,\bar\theta^2$ and higher) are zero.
The $\theta\bar\theta$ component of $J$ is a conserved vector current.
One gauges the symmetry by introducing a (real) vector multiplet $V$,
whose components are sources for a multiplet of currents.
There is a linear coupling in the microscopic action
\begin{equation}\label{JVcoupling}
\int \rd^4 x\,\rd^2\theta\,\rd^2\bar\theta\,VJ
\end{equation}
and current conservation \eqref{conservation} is translated to the
gauge invariance of the generating functional  for the transformation
\begin{equation}\label{deltaV}
V'=V+{\ri\over2}(\Lambda-\Lambda^{\dagger})\,,\qquad \bar D_{\dot\a}\Lambda=0
\end{equation}
where $\Lambda$ is a chiral scalar superfield.
$V=V^\dagger$ has the expansion
\begin{equation}
\begin{aligned}
V(x,\theta,\bar\theta)&
=C+\ri\,\theta\chi-\ri\,\bar\theta\bar\chi+{\ri\over2}\theta^2 M-{\ri\over2}\bar\theta^2\bar M-\theta\sigma^m\bar\theta\, v_m\\
&+\ri\,\theta^2\,\bar\theta\left(\bar\lambda+{\ri\over2}\bar\sigma^m\partial_m\chi\right)
-\ri\,\bar\theta^2\,\theta\left(\lambda+{\ri\over2}\sigma^m\partial_m\bar\chi\right)
+{1\over2}\theta^2\,\bar\theta^2\left(D+{1\over2}\square C\right)\,.
\end{aligned}
\end{equation}
Here $C$ and $D$ are real scalars, while $M$ is complex; $v_m$ is a real vector and $\chi$ and $\l$ are
Weyl spinors.
In this case the gauge group $\mathcal{G}$ is simply the additive group of chiral scalar superfields as defined by \eqref{deltaV}.

The generating functional $\Gamma[V]$ is not gauge invariant, i.e. there is an anomaly given by
\begin{equation}\label{flavour_anomaly}
\delta_\Lambda\Gamma[V]=\ri\int \rd^4 x\,\rd^2\theta\,{\Lambda}\, W^\a W_\a+\hbox{h.c.}
\end{equation}
for an infinitesimal $\Lambda$ or, equivalently,
\begin{equation}
\bar D^2\langle J\rangle=8\,W^\a W_\a\qquad\hbox{where}\qquad
\langle J\rangle={\delta\over\delta V}\Gamma[V]\,.
\end{equation}
Here $W_\a=-{1\over4}\bar D^2\,D_\a V$ is the (chiral) gauge-invariant field strength and we put the strength of the anomaly to $1$ for convenience.

We now want to study the anomaly in the WZ gauge where by a partial gauge fixing the  lower components
$C,\chi$ and $M$ are gauged to zero, i.e.
\begin{equation}
\begin{aligned}
V(x,\theta,\bar\theta)|_{\rm WZ}&
=-\theta\sigma^m\bar\theta\, v_m
+\ri\,\theta^2\,\bar\theta\bar\lambda
-\ri\,\bar\theta^2\,\theta\lambda
+{1\over2}\theta^2\,\bar\theta^2\,D\equiv\bar V
\end{aligned}
\end{equation}
This is achieved by making a gauge transformation defined by the special choice of $\Lambda$
\begin{equation}\label{G0}
\Lambda_0= \ri\,C(y)-2\,\theta\chi(y)-\theta^2M(y)\,,
\end{equation}
where $y\equiv x+\ri\theta\sigma\bar{\theta}$.
We have therefore the relation
\begin{equation}\label{AGA0}
V(x,\theta,\bar\theta)|_{\rm WZ}=V(x,\theta,\bar{\theta})+{\ri\over2}(\Lambda_0-\Lambda_{0}^{\dagger})\,.
\end{equation}
From \eqref{AGA0} it is clear that $\Lambda_0$ plays the role of $G_0(A)$ in our general discussion.
In particular the gauge fixing is purely algebraic.
However, as is obvious from \eqref{G0} and \eqref{AGA0}, this gauge is inconsistent with supersymmetry:
while being a chiral superfield,
i.e. $\bar D_{\dot\a}\Lambda_0=0$, the supersymmetry transformations of its components
are those of the components of a real superfield and not of a
scalar chiral superfield.\footnote{The conflict with supersymmetry exists for any partial WZ gauge,
where we transform away only $C$ or $C$ and $\chi$.}

After fixing the WZ gauge, the residual, non-algebraic,  gauge transformations are generated by
$\Lambda=\a$ with $\a$ real. Under these transformations only $v_m$ transforms: $v_m\to v_m+\partial_m\a$.
The anomaly in the WZ gauge corresponds to the standard chiral $U(1)$ gauge anomaly
\begin{equation}\label{anom}
\delta_{\a} \Gamma_{\rm WZ}[\bar V]=\int \rd^4x\, \a\, v_{mn}\tilde v^{mn}
\end{equation}
where $v_{mn}$ is  the field strength of $v_{m}$. Up to the gauge variation of the local term
$2\int \rd^4 x\,v_m\,\bar\lambda\sigma^m\lambda$, the anomaly in \eqref{anom} is the restriction
of the general anomaly \eqref{flavour_anomaly} for $\Lambda=\a$.
The cohomology in the WZ gauge is completely represented by the restriction of the action to
configurations $\bar{V}$ in the WZ gauge. A generic effective action
calculated in  the WZ gauge is therefore the restricted action $\Gamma[V=\bar{V}]$ modulo local terms.
We will therefore be able to follow our general treatment where $\bar{V}$ corresponds to $\bar{A}$ , etc.

We now want to find a counterterm which, when added to
the superspace functional $\Gamma[V]$, will reproduce
the WZ gauge results.
Finding the WZ functional is trivial due to the abelian nature of the gauge transformation. We have
\begin{equation}\label{abel}
\Gamma[V']=\Gamma[V]+\ri\int \rd^4 x\,\rd^2\theta\Lambda\, W^\a W_\a+\hbox{h.c.}
\end{equation}
for any finite $\Lambda$, $V'$ being the gauge transform of $V$.
Then the counterterm $\mathcal{C}$ following \eqref{uppp} is
\begin{equation}\label{counter}
\mathcal{C}\equiv \ri \int \rd^4 x\, \rd^2\theta\,\Lambda_0\,W^\a W_\a+\hbox{h.c.}
\end{equation}
If we define
\begin{equation}
\tilde\Gamma[V]=\Gamma[V]+{\cal C}[V]
\end{equation}
where ${\cal C}$ plays the role of $W(A;G_0(A))$ in \eqref{uppp},
we have
\begin{equation}\label{check}
\tilde\Gamma[V+\textstyle{\ri\over2}(\Lambda_0+\Lambda_0^\dagger)]=\tilde\Gamma[V]
\end{equation}
and the counterterm corrected generating functional has the following properties:

a) its anomaly reproduces \eqref{anom}

b) it depends only on the $V$-components in the WZ gauge

\noindent

It follows from the latter property that $\tilde\Gamma[V]=\Gamma[\bar{V}]$.
The meaning of this definition  is that for every $V$
we identify its unique $\bar V$ representative which can be reached from $V$ by an ultralocal gauge
transformation  and then $\tilde\Gamma[V]$ is defined as $\Gamma[\bar V]$. We stress that
$\bar V$ is not an independent variable but it is determined  by $V$.
Moreover all $V$-configurations of the form $\bar{V}+{\ri\over2}(\Lambda-\Lambda^{\dagger})$,
where $\Lambda$ is a chiral superfield with purely imaginary lowest component, will have the same representative
and therefore the same value of $\tilde\Gamma$.
$\tilde\Gamma[V]$ is not supersymmetric because, as we have remarked before, the counterterm explicitly
breaks supersymmetry due to the ``wrong" transformation properties of $\Lambda_0$.

We want to calculate the action of supersymmetry on the generating functional in two ways, which should agree:

a) in the WZ gauge  where the generating functional is $\Gamma[\bar{V}]$;

b) on $\tilde\Gamma$.

We start with a):
The  supersymmetry transformation does not preserve the
gauge choice. To correct for this one has to accompany it by a compensating
{\it field dependent} gauge transformation:
\begin{equation}
\hat\delta_\epsilon=\delta_\epsilon+\delta_{\Lambda(\epsilon)}
\end{equation}
where
\begin{equation}
\delta_\epsilon\Psi=\big[\epsilon Q+\bar\epsilon\bar Q,\Psi \big]
\end{equation}
for any superfield $\Psi$ and
\begin{equation}
\Lambda(\epsilon)=-2\ri\,\theta\sigma^m\bar\epsilon\,v_m(y)-2\,\theta^2\,\bar\epsilon\bar\lambda(y)\,.
\end{equation}
The compensating gauge transformation brings the components $\chi,M$ of $V$, which
are reintroduced by $\delta_\epsilon$,  back to zero.
Therefore the supersymmetry action on the generating functional, $\bar{V}$ being an independent variable, is
\begin{equation}\label{SUSY}
\hat{\delta_{\epsilon}} \Gamma[\bar{V}]=\delta_{\Lambda({\epsilon})}\Gamma[\bar {V}]
=\ri\int \rd^4 x\,\rd^2\theta\Lambda({\epsilon})\, W^\a(\bar{V}) W_\a(\bar{V})+\hbox{h.c.}
\end{equation}

We now proceed to b):
The basic idea is to make an independent  supersymmetry variation on $V$, then identify the new
``representative" $\bar{V}$ and evaluate $\Gamma$ for the new representative. It is essential that  the new $\bar{V}$
is not a supersymmetry variation of the old representative $\bar{V}$.

We start with a supersymmetry variation of $\bar{V}$ itself, $\delta_{\epsilon}\bar{V}$. We write this as
\begin{equation}\label{triv}
\delta_{\epsilon}\bar{V}=\hat{\delta_{\epsilon}}\bar{V}-\delta_{\Lambda({\epsilon})}\bar{V}\,.
\end{equation}
Since $\Lambda({\epsilon})$ defined above is ultralocal, this shows
that the representative on the orbit of $\bar{V}+\delta_{\epsilon}\bar{V}$ is $\bar{V}+\hat{\delta_{\epsilon}}\bar {V}$,
the corresponding $G_0$ being $-\Lambda({\epsilon})$.
Therefore
\begin{equation}\label{count}
\delta_{\epsilon}\tilde\Gamma[\bar{V}]\equiv\delta_{\epsilon}\mathcal{C}[\bar{V}]=\hat{\delta_{\epsilon}}\Gamma[\bar{V}]\,.
\end{equation}
From \eqref{count} it is apparent that what looked in the WZ gauge as an anomaly became in the new scheme the variation
of the local counterterm $\mathcal{C}$, which violates explicitly supersymmetry for the reasons discussed above.

By adding the additional counterterm (cf.below \eqref{anom})
\begin{equation}
{\cal C}_1=2\int \rd^4 x\,\bar\l\,\bar\s^m\l\,v_m
\end{equation}
which involves only fields which are present in the WZ gauge and using \eqref{SUSY}, one finds:
\begin{equation}
\delta_\e(\mathcal{C}+{\cal C}_1)[\bar{V}]=2\int \rd^4 x\Big( 3\,\ri\,\e\l\,\bar\l^2
+\ri\,\e\s_q\bar\l\,\e^{mnpq}v_{mn}\,v_p
+\hbox{h.c.}\Big)\,.
\end{equation}
This agrees with the `SUSY-anomaly' of refs. \cite{Nair,Guadagnini,Closset}.

Equation \eqref{count} can be generalized for arbitrary values of $V$, but one gets an additional contribution related to
a genuine gauge transformation component in the supersymmetry transformation of $v_m$.

We end this section with a comment on the correlators which can be derived from $\Gamma$ and
$\tilde\Gamma$ as functional derivatives w.r.t. the sources.
From the former, which depends on all components of the vector superfield $V$,
one derives the correlators involving all components of the current $J$. There are purely local terms contained
in $\Gamma[V]$ which are linear in the sources $(C,\chi,M)$. They can be read off from the component
expansion of the counterterm ${\cal C}$ and the fact that $\tilde\Gamma[V]$ is independent of them.
The operators of the microscopic theory, to which the components of $V$ couple, are read off from the $JV$ coupling.
Those coupling to $(C,\chi,M)$ are ultralocal null operators.
For instance, $\Gamma[V]$ contains the term $M\lambda^2$; e.g. in the free massless WZ model, one finds:
$M$ couples to $A\bar F$ and $\lambda$ couples to $\psi\bar A$,
where $(A,\psi,F)$ are the components of a chiral multiplet. One easily verifies that the three point function
$\langle A\bar F\,\psi\bar A\,\psi\bar A\rangle$ is purely local and, e.g. in Pauli-Villars
regularization, only the regulator field contributes.
The terms in $\Gamma[V]$ which depend on
$\psi$ and $M$ can be analysed similarly. They also correspond to purely local three-point functions
with one operator insertion sourced by $\chi$ or $M$.

In contrast to this, $\tilde\Gamma$ does not carry information about the correlators of the redundant operators, but
they are needed for the supersymmetric Ward identities to be satisfied. To restore them, the local correlators
derived from the counterterm ${\cal C}$ have to be added by hand. They contain components of $J$
whose sources are absent in $\tilde\Gamma$, but they can be recovered as explained in general terms
in the introduction and explicitely for the $U(1)$ flavour current in this section.


\section{Super-Weyl anomalies}

In the following we are interested in the quantum aspects of superconformal field theories
coupled to  supergravity. There exist two powerful superspace formulations
for $\cN=1$ conformal supergravity \cite{KTvN1,KTvN2}:
(i) the U(1) superspace of \cite{Howe} (see \cite{GGRS} for a review);
and (ii) the conformal superspace developed in \cite{Butter}.\footnote{The
conformal superspace approach \cite{Butter} is a master formulation for conformal supergravity.
All other off-shell formulations, including the superconformal tensor calculus
\cite{KTvN1,KTvN2} (see, e.g.,  \cite{FT-conformal} for a review),
can be obtained from conformal superspace by partially fixing the gauge freedom.}
However the simplest and most economical  approach to describe $\cN=1$ conformal supergravity in superspace
is to make use of the Grimm-Wess-Zumino geometry \cite{GWZ}, which underlies the Wess-Zumino formulation for
old minimal supergravity \cite{WZ} (see \cite{WB} for a review)
developed independently in \cite{old1,old2}.
In order to formulate conformal supergravity, the gauge
group of old minimal supergravity \cite{WB} has to be extended to include the
super-Weyl transformations, originally introduced in \cite{HT}.
The specific feature of superconformal field theories is that  they are invariant under
arbitrary  super-Weyl transformations.

The supergravity multiplet is described by covariant derivatives
$\cD_A =(\cD_a, \cD_\a ,\bar \cD^\ad)$, eq. \eqref{A.111},
such that the torsion and curvature tensors obey nontrivial constraints \cite{WZ}. 
These constraints
were solved by Siegel \cite{Siegel78} in terms of two unconstrained prepotentials,
a real axial vector $H^m (x,\q, \bar \q)$ and a chiral density $\vf (x,\q)$.
The former is equivalent to the gravitational superfield introduced by Ogievetsky and
Sokatchev \cite{OS}. The latter determines the integration measure $\cE$ of chiral
subspace, $\cE = \vf^3$.

A finite super-Weyl transformation acts on $H^m$ and $\vf$ by the rule
\cite{Siegel:SC}
\bea
H^m \to H^m~, \qquad \vf \to \re^\S \vf~,
\label{2.1}
\eea
with $\S$ a covariantly  chiral scalar.
This transformation law implies that locally superconformal field theories
do not couple to $\vf$ at the classical level, 
since all dependence on $\vf$ can be absorbed into matter supermultiplets.

Another fundamental symmetry group acting on $H^m$ and $\vf$ is the supergroup of
``holomorphic coordinate transformations"
(``$\lambda$-transformations" in the following):
\begin{equation}\label{finite}
y^{\prime m}=f^m(y,\theta)\,,\quad\bar y^{\prime m}=\bar f^m(\bar y,\bar\theta)\,,\quad
\theta'^{\a}=f^{\a}(y,\theta)\,,\quad
\bar{\theta}^{\prime\dot{\a}}=\bar{f}^{\dot{\a}}(\bar y,\bar{\theta})\,.
\end{equation}	
Infinitesimal holomorphic coordinate transformations are parametrized by chiral
superfields $\lambda^{m}(y,\theta) ,\lambda^{\a}(y,\theta)$ and their complex conjugate anti-chiral fields
$\bar\lambda^m(\bar y,\bar\theta),\bar{\lambda}^{\dot{\a}}(\bar y,\bar{\theta})$, e.g.
for infinitesimal transformations
\begin{equation}\label{infi}
f^m(y,\theta)=y^m-\lambda^{m}(y,\theta)\,,\qquad
f^{\a}(y,\theta)=\theta^{\a}-\lambda^{\a}(y,\theta)\,.
\end{equation}
The transformations of the real coordinates $x^m$ are defined in an implicit fashion by
\begin{equation}\label{messl}
\begin{aligned}
x^m&\to x^{\prime m}={1\over2} f^m(x+\ri\,H,\theta)+{1\over2}\bar{f}^m(x-\ri\,H,\bar\theta)\,,\\
\noalign{\vskip.2cm}
\theta^\a&\to\theta^{\prime\a}=f^\a(x+\ri\,H,\theta)\,,\qquad
\bar{\theta}'^{\dot\a}\to\bar\theta^{\dot\a}=\bar{f}^{\dot\a}(x-\ri\,H,\bar\theta)\,.
\end{aligned}
\end{equation}
The transformations of the gauge fields $H^m$ and $\vf$ under $\lambda$-transformations are
\begin{equation}\label{messH}
H^{\prime m}(x',\theta',\bar\theta')=
-{\ri\over2}f^m(x+\ri\,H,\theta)+{\ri\over2}\bar{f}^m(x-\ri\,H,\bar\theta)
\end{equation}
and
\begin{equation}\label{messphi}
\varphi'(y',\theta')=\left[{\rm Ber}\left({\partial(y,\theta)\over\partial(y',\theta')}\right)\right]^{1/3}\varphi(y,\theta)
\end{equation}
respectively.  `Ber' denotes the Berezinian also known as the superdeterminant.
The infinitesimal versions of these transformations are
\begin{subequations}
\bea
\d_\l H^m &=& \frac{\ri}{2} \l^m (x +\ri H , \q) -\frac{\ri}{2}\bar  \l^m (x -\ri H , \bar \q) \non \\
&&
+ \bigg( \hf  \l^n (x +\ri H , \q) \pa_n + \l^\a (x +\ri H , \q) \pa_\a +{\rm c.c.} \bigg)
 H^m~,
\\
\d_\l \vf &=& (\l^m \pa_m +\l^\a \pa_\a ) \vf + \frac 13 ( \pa_m \l^m -\pa_\a \l^\a)\vf ~.
\eea
\end{subequations}

It is worth remarking that in the case of Minkowski superspace
the prepotentials $H^m$ and $\vf$ can be chosen in the form
\begin{subequations}\label{flat}
\bea
H^m &=& \q \s^a \bar \q \, \d_a{}^m~, \label{flat-H}\\
\vf &=& 1~, \label{flat-vf}
\eea
\end{subequations}
by partially fixing the $\l$ gauge freedom.
The rigid superconformal transformations of Minkowski superspace
are those $\l$-transformations which preserve $H^m$ given by \eqref{flat-H}.
They are (see e.g. \cite{BK})
\begin{subequations} \label{rigid-superconformal}
\bea
\l^{a}(y,\theta) &=& {a}^{a}  + \s y^{a} + K^a{}_b y^{b}
+y^a f^2 - 2f^a f_b y^b
+2{\rm i}\, \q \s^a \bar \e
-  \q \s^a \tilde{\s}_b \eta y^b ~, \\
\l^\a(y,\theta) &=& \e^\a - \frac{\ri}{2} (\bar \eta \tilde \s_b)^\a y^b
+ \hf \big(\s  +\ri  \r \big)  \q^\a - K^\a{}_\b \q^\b
+f^a y^b (\q \s_a \tilde \s_b)^\a
+ \eta^\a \q^2 ~,
\eea
\end{subequations}
with all the parameters being constant.
Here the real scalar parameters $\s$ and $\r$ generate scale and $R$-symmetry transformations.
The real vectors $a^a$ and $f^a$ correspond to the spacetime translations and special conformal transformations, respectively,
while the real antisymmetric parameter $K_{ab}$
generates the Lorentz transformations. Finally,
the spinor parameters $(\e^\a, \bar \e^\ad)$ and $(\eta^\a , \bar \eta^\ad)$
generate the $Q$ and $S$ supersymmetry transformations, respectively.
The isometries of Minkowski superspace
are those $\l$-transformations which preserve both $H^m$ and $\vf$
 given by eqs. \eqref{flat-H} and \eqref{flat-vf}.
 They are obtained from
 \eqref{rigid-superconformal} by switching off the parameters $\s$, $\r$,
 $f^a$ and $\eta^\a$.

Let $S [\c, H, \vf , \bar \vf] $ be the action of matter superfields $\c$
(with suppressed indices) coupled
to  the supergravity sources.  The coupling should be invariant under
 $\lambda$-transformations.
All information about the coupling of matter to supergravity
is encoded in two tensor superfields, the supercurrent $J_a =\bar J_a$
and the supertrace $T$, which originate as
covariantised variational derivatives of $S [\c, H, \vf , \bar \vf] $
with respect to the supergravity prepotentials.
If the matter and source superfields are given small disturbances, the action
varies as
\bea \label{2.2}
\d S =  \int \rd^4 x \rd^2 \q \rd^2 \bar \q \,E \, \D H^{\a\ad} J_{\a\ad}
+\bigg\{\int \rd^4 x \rd^2 \q \,\cE \, \d \ln \vf \,T +{\rm c.c.} \bigg\}
+\int {\delta S\over\delta\chi}\delta\chi~.
\eea
Here $E^{-1} =  {\rm Ber} (E_A{}^M)$  is the full superspace  measure,
$\cE= \vf^3$  the chiral measure and
$\D/ \D H^{\a\ad}$ denotes a covariantised variational derivative with respect to the
gravitational superfield \cite{GrisaruSiegel}. By construction, the supertrace is covariantly chiral, $\bar \cD_\ad T=0$.

If the matter supermultiplets obey their equations of motion, $\d S/ \d \c=0$,
the condition that the matter action is invariant under $\lambda$-transformations  is expressed
as the conservation equation
\bea\label{diver}
\bar \cD^\ad J_{\a\ad} = \frac 13 \cD_\a T ~.
\eea
The supercurrent
$J_{\a\ad}$ reduces to the Ferrara-Zumino
multiplet \cite{FZ75} when the sources are put to zero, eq. \eqref{flat}.

In a superconformal field theory, the super-Weyl transformation \eqref{2.1} is
accompanied by a local rescaling
of the matter supermultiplets of the form $\c \to \c' = \re^{-d_{+} \S - d_{-} \bar \S}\c$, for some parameters $d_{\pm} $,
such that the action is super-Weyl invariant,
\bea
\d_\S S =
\int \rd^4 x\, \rd^2 \q \,\cE \, \S\, T + {\rm c.c.} +\int \d_\S\c  \,\frac{\d S}{\d \c} =0~.
\eea
This implies that the classical supertrace vanishes,
\bea\label{trace}
T=0~,
\eea
on the mass shell.
The two conservation equations \eqref{diver} and \eqref{trace} following from the two classes of symmetries cannot be both
implemented in the quantum theory, leading to the superconformal anomalies.

We now proceed to a detailed discussion of the quantum theory.
Integrating out the matter supermultiplets results in an effective action $\G[H, \vf , \bar \vf]$.
The effective action $\Gamma$ does not respect both the super-Weyl and $\lambda$-transformation symmetries,
i.e. an anomaly appears. In this paper we will use as a basic starting point in superspace the
``superspace scheme" where $\lambda$-transformations  are preserved.\footnote{In the non-supersymmetric
setting this would correspond to the scheme where diffeomorphism invariance is kept and
Weyl symmetry is sacrificed; cf. Appendix B.}

Invariance of the effective action under $\l$-transformations is encoded in the conservation equation
\bea\label{cons}
\bar \cD^\ad \langle J_{\a\ad}  \rangle = \frac 13 \cD_\a \langle T  \rangle~.
\eea
On the other hand in the quantum theory \eqref{trace} is violated.
Here the quantum supertrace
is defined by
\bea\label{qtrace}
\d_\S  \G [ H,  \vf ,  \bar \vf] & &=
\int \rd^4 x\, \rd^2 \q \,\cE \, \S\, \langle T \rangle + {\rm c.c.} ~\label{defT}\\
\noalign{\noindent where}
\d_\S H^m=0 & &~,\qquad \d_\S \vf = \S\, \vf\,.
\label{supertrace}
\eea
The appearance of  a super-Weyl anomaly in the quantum theory, i.e. the violation of \eqref{trace},
means that the effective action acquires a dependence on the chiral prepotential $\vf$ and its conjugate $\bar \vf$,
unlike the classical action of a superconformal field theory. This dependence is
``cohomologically  nontrivial'' in the sense that  by adding
local counterterms
the effective action cannot be made independent of $\vf$ and $\bar \vf$
without spoiling the invariance under $\lambda$-transformations. In the next section we will analyze in
detail the subset of $\lambda$-transformations  which are incompatible with the vanishing of the quantum supertrace.

According to the cohomological analysis of \cite{BPT} and explicit supergraph calculations\footnote{The work described in \cite{BK86} was completed in 1984
(the same year as \cite{BPT}) 
 but then it took over a year to obtain 
the KGB clearance required for publication in the West.}
 for the scalar
and vector supermultiplets \cite{BK86},
the general  form of $\langle T\rangle $
in classically super-Weyl invariant theories is
\bea
 \langle T\rangle= 2(c-a)  W^{\a\b\g} W_{\a\b\g}
 +\hf a  (\bar \cD^2 - 4R ) (G^aG_a +2R \bar R)
 ~,
\label{s-Weyl-anomaly}
\eea
modulo cohomologically trivial contributions. Here {$a$ and $c$ are two numerical coefficients whose values
depend on the microscopic superconformal field theory.\footnote{The general
structure of
the super-Weyl anomaly, eq. \eqref{s-Weyl-anomaly}, can be extracted from the earlier work of McArthur \cite{McA}.}
Equation \eqref{s-Weyl-anomaly} is equivalent to the fact that the super-Weyl variation of the effective action is
($\rd^{4|4}z = \rd^4 x \rd^2 \q \rd^2 \bar \q$)
\bea
\d_\S \G  [H, \vf , \bar \vf ]= 2(c&-&a) \int \rd^4 x \rd^2 \q \,\cE \,\S W^{\a\b\g} W_{\a\b\g} +{\rm c.c.}
\non \\
&-&2a  \int \rd^{4|4}z \,E \, (\S+\bar \S)(G^aG_a +2R \bar R)~.
\label{2.4}
\eea

We are interested in
a local action $K=K [H, \vf, \bar \vf; \O , \bar \O] $ (also called ``WZ action" in the following)
that gives the transformation of $\Gamma$ under a finite super-Weyl transformation $\O\equiv \exp{\Sigma}$,
where $\O$ is a covariantly chiral scalar superfield:
\begin{equation}\label{WZA}
K\equiv \G [H_{\O},\vf_{\O},\bar{\vf}_{\O}]-\G[H,\vf,\bar{\vf}]
\end{equation}
with
\begin{equation}\label{def}
H_{\O}=H\,,   ~~    \vf_{\O}=\O\vf\,,  ~~     \bar{\vf}_{\O}=\bar{\O}\bar{\vf}
\end{equation}
The required action
was constructed in
\cite{SchT} by integrating the anomaly to finite transformations with parameter $\Omega$
and has the form
\bea\label{WZ}
&& K [H, \vf, \bar \vf; \O , \bar \O]
=  2(c-a) \int \rd^4 x \rd^2 \q \,\cE \,\ln \O\, W^{\a\b\g} W_{\a\b\g} +{\rm c.c.} \non \\
&& \qquad  -2\,a  \int \rd^{4|4}z \,E \,
\bigg\{\ln (\O \bar \O )
(G^aG_a +2R \bar R)
-\hf G^{\a\ad} \bar \cD_\ad \ln \bar\O\, \cD_\a \ln \O \non \\
&& \qquad\qquad -\frac 14 \Big( R (\cD \ln \O)^2 + \bar R (\bar \cD \ln \bar \O)^2 \Big)
 + \frac{1}{16} (\cD \ln \O)^2  (\bar \cD \ln \bar \O)^2
\non \\
&& \qquad\qquad\qquad+\frac{\ri}{4} \cD^{\a\ad} (\ln \bar \O - \ln \O)  \bar \cD_\ad \ln \bar \O\, \cD_\a \ln \O
 \bigg\}~.
\label{3.8}
\eea
An alternative derivation is presented in  Appendix \ref{AppendixB}.
Under the infinitesimal super-Weyl transformation \eqref{supertrace}
accompanied with 
\bea
\d_\S \O = - \S \O~,
\label{3.222}
\eea
the functional \eqref{3.8} varies as 
$
\d_\S K [H, \vf, \bar \vf; \O , \bar \O] =- \d_\S \G  [H, \vf , \bar \vf ]$.

In addition to the above equation defining the super-Weyl transformation of the ``superspace scheme" effective action,
we should also give  the transformation under $\lambda$-transformations which is simply
\bea\label{supl}
\delta_{\lambda}\G[H,\vf,\bar{\vf}]= 0\,,
\eea
where the transformations of the arguments are given in \eqref{messH} and \eqref{messphi}.

Given the fact that $\vf$ is needed  just in the presence of the super-Weyl anomaly,
one wonders how unique the completion of $H$ by this additional degree of freedom is.
As an extension of the Weyl-invariant formulation for gravity \cite{Deser,Zumino},
every off-shell supergravity theory can be realised as a super-Weyl invariant coupling
of conformal supergravity to a compensating supermultiplet, see e.g.
\cite{SG,deWR,FGKV,GGRS}.
In such a setting, any supergravity-matter system is described by a super-Weyl invariant
action functional. Different off-shell supergravity theories correspond to different compensating
supermultiplets.
Locally superconformal theories are independent of any compensator.
In the case of a classically superconformal theory,
the presence of a super-Weyl anomaly at the quantum level is equivalent to a nontrivial dependence of the effective
action on the compensating supermultiplet,
as advocated in \cite{K13}. The compensating super-Weyl invariance  is not anomalous
\cite{deWG}.

In the case of old minimal supergravity \cite{WZ,old1,old2}, the compensator
is a nowhere vanishing covariantly chiral scalar $S_0$, $\bar \cD_\ad S_0 =0$,
with the super-Weyl transformation
\bea
S_0 \to \re^{-\S} S_0~.
\eea
In the superconformal setting, the effective action $ \G [H, \vf , \bar \vf ]$ is replaced by the
following super-Weyl invariant functional
\bea
\G [H, \vf, \bar \vf, S_0 , \bar S_0] &=&  \G [H, \vf , \bar \vf ]
+ K[H,\vf,\bar\vf,S_0,\bar S_0]\,.
\label{SWIEE}
\eea
In the compensator approach, the super-Weyl anomaly is manifested in the dependence of
$\G [H, \vf, \bar \vf, S_0 , \bar S_0] $ on $S_0$ and its conjugate $\bar S_0$. Choosing the super-Weyl gauge
$S_0 =1$ reduces $ \G [H, \vf, \bar \vf, S_0 , \bar S_0] $ to the original effective action,
$\G [H, \vf, \bar \vf] $.

In the framework of the new minimal formulation for $\cN=1$ supergravity
\cite{SW1,SW2}, the compensator is a covariantly
linear supermultiplet\footnote{The linear compensator \cite{deWR}
is described by a tensor multiplet
\cite{Siegel-tensor} such that its field strength $L$ is
nowhere vanishing.} $ L$ constrained by
\bea
(\bar \cD^2 -4R) { L}  =0~, \qquad \bar { L}= { L}~.
\eea
Its super-Weyl transformation is uniquely fixed by these constraints to be \cite{BK}
\bea
L \to \re^{-\S -\bar \S} L~.
\eea
Unlike the effective action \eqref{SWIEE} constructed using the chiral compensator $S_0$,
 it appears that there is no way to complete
$ \G [H, \vf , \bar \vf ]$ to a super-Weyl invariant functional by adding local structures depending
only on the linear compensator, in addition to $H^m,\varphi$ and $\bar\varphi$.
This is analogous to the non-minimal formulation for $\cN=1$ supergravity \cite{non-min,SG},
for which the compensator is a complex linear supermultiplet $\U$, only constrained by
\bea
(\bar \cD^2 -4R) \U  =0~.
\eea
Under super-Weyl transformation it transforms as \cite{BK}
\bea
\U \to \exp \Big( \frac{3n-1}{3n+1} \S   -\bar \S \Big) \U~,
\eea
with $n \neq  -1/3 , 0$ a real parameter.
It does not seem to be possible to complete
$ \G [H, \vf , \bar \vf ]$ to a super-Weyl invariant functional by adding local structures depending only on $\U$
and its conjugate $\bar \U$.
These conclusions agree with the old analysis of \cite{GGS}
which established the incompatibility of  the new minimal and non-minimal supergravity
formulations with the existence of
local super-Weyl anomalies.
In this sense indeed the $H$, $\vf$ setup summarised above is unique and  we will formulate all
our further discussion in this framework.


\section{The minimal Q-supersymmetric scheme}

We will first try in superspace to find a scheme in which the anomaly in super-Weyl transformations is shifted to
$\lambda$-transformations. This will allow us to identify those $\lambda$-transformations which are necessarily anomalous
in such a scheme.

Starting with the standard scheme $\G$, we want to reach the configuration $H,\vf=1,\bar{\vf}=1 $.
This can be achieved by doing a super-Weyl transformation with $\O=\vf^{-1}$ .
As the super-Weyl transformation is anomalous, in order to have
$\G$ at the new configuration we have to use \eqref{WZA}  and \eqref{WZ} i.e.
\begin{equation}\label{H11}
\G[H,1,1]=\G[H,\vf,\bar{\vf}]+K [H,\vf,\bar{\vf},\vf^{-1},\bar{\vf}^{-1}]\,.
\end{equation}
The r.h.s. of the equation is $\vf$-independent.
Therefore $\G[H,1,1]$ is related to $\G[H,\vf,\bar{\vf}]$
by a local counterterm and we can  define
a new scheme with a new generating functional  $\tilde{\G}$ by
\begin{equation}\label{tilde}
\tilde{\G}[H]\equiv \G[H,1,1]\,.
\end{equation}
In the new scheme $\tilde{\G}$ is independent of $\vf$ and therefore the super-Weyl anomaly vanishes, i.e.
\begin{equation}\label{ntrace}
\tilde{T}=0\,.
\end{equation}
The variation under finite $\lambda$ transformations is not difficult to calculate:
\bea
\delta_{\lambda}\tilde \G & \equiv & \G[ H_{\lambda},1,1]-\G[H,1,1] \non \\
&=& \Big[\G[ H_{\lambda},B,\bar B]-\G[H,1,1]\Big]
- \Big[\G[ H_{\lambda},B,\bar B]-\G[ H_{\lambda},1,1]\Big]~,
\label{lamvar}
\eea
where $B$ denotes the weight factor  in  \eqref{messphi},
\bea
B = \left[{\rm Ber}\left({\partial(y,\theta)\over\partial(y',\theta')}\right)\right]^{1/3}~,
\eea
associated with the  $\lambda$-transformation.
The expressions in brackets in the
second line of
\eqref{lamvar} are variations for the standard scheme: the first is zero
due to $\lambda$-invariance of $\Gamma$,  while
the second is given by \eqref{WZ} for $\vf =1$ and $\O=B$.
In the infinitesimal case
\bea
B = 1 + \frac 13 \big( \pa_m \l^m - \pa_\a \l^\a\big)  \equiv 1+\S(\l)
\label{B}
\eea
and therefore the anomaly becomes
\bea\label{lan}
\delta_{\lambda}\tilde\G[H]
&=& -\int \rd^4 x \rd^2 \q  \,\S(\l) \langle T\rangle +{\rm c.c.}~,
\eea
where $\langle T\rangle $ is evaluated at $\vf=1$.
In particular we see  that only $\lambda$-transformations with $B\neq 1$ are anomalous.

We will discuss first the Ward identities associated with the non-anomalous symmetries in this scheme.
In addition to \eqref{ntrace} the infinitesimal form of \eqref{lan}
may be shown to give
\bea\label{ncons}
\bar \cD^\ad \langle \tilde{ J}_{\a\ad}  \rangle = \frac 13 \cD_\a \langle T  \rangle~.
\eea
with $T$ given by \eqref{s-Weyl-anomaly}, evaluated in the configuration $(H,1,1)$.
We remark that the appearance of $T$ and not of $\tilde T$ in the r.h.s. of \eqref{ncons} shows that in this new scheme
the $\lambda$-transformations are anomalous. On the other hand the ``improvable" form of
\eqref{ncons} is related to the fact that the subgroup of $\lambda$-transformations with unit Berezinian
remains non-anomalous.

Once the symmetries of the ``minimal $Q$-symmetric" scheme defined by the addition of  the local counterterm \eqref{H11}
are understood, we could  study the detailed properties of $\tilde\Gamma$ in the ``ultralocal" gauge it defines.
There is a one-to-one correspondence between
the ultralocal gauge condition, given by the field dependent gauge group functional, $G_0(A)$ in the notation of
the introduction and the scheme in which the gauge transformations $G_0(A)$ is anomaly free.
In this example we started with the scheme and  we identified as the anomaly free direction
the $\lambda$-transformations with unit Berezinian and $\vf,\bar{\vf}$ fixed to $1$. Therefore
the ultralocal gauge transformations correspond to $\lambda$-transformations $f_0(H) $ which brings $H$ to the form:
\bea\label{H-WZgauge2}
\tilde{H}^m (\q, \bar \q) &=& \frac{\ri}{2} \q^2 S^m - \frac{\ri}{2} \bar \q^2 \bar S^m
+ \q \s^a \bar \q e_a{}^m + \ri \,\bar \q^2 \q^\a  \J^{m}{}_{ \a}
-\ri \,\q^2  \bar \q_\ad  \bar \J^{m \ad}  \non \\
&&+  \q^2 \bar \q^2 \Big( A^m -\frac{\ri}{4 } \Big[ S^n \pa_n (\bar S^m)
- \bar S^n \pa_n (S^m) \Big] \Big)~,
\eea
while
\begin{equation}\label{vf-WZgauge2}
\vf=1 ~, \qquad \bar{\vf}=1\,.
\end{equation}
is reached by a super-Weyl transformation with $\Sigma=-\log\vf$ under which $H^m$ is invariant.
The relations \eqref{H-WZgauge2} and \eqref{vf-WZgauge2}
define a  WZ gauge in old minimal supergravity.

We will not need the explicit form of $f_{0}(H)$. Since the transformation is
non-anomalous,  the $H$ configuration given by \eqref{H-WZgauge2} is a convenient labelling of the gauge orbit.
By construction the generating functional in this scheme is independent of the three lowest components
$H$, i.e. those which do appear in \eqref{H-WZgauge2} as they were gauged away:
\begin{equation}\label{rest-H}
\tilde{\Gamma}[H]=\tilde{\Gamma}[H=\tilde{H}] \,,
\end{equation}
The effective action $\tilde\Gamma[H]$ is invariant under those gauge transformations
which are used to arrive at the WZ gauge
conditions \eqref{H-WZgauge2} and \eqref{vf-WZgauge2}, namely
volume preserving $\l$-transformations and arbitrary super-Weyl transformations.

We now study the remaining symmetries of the generating functional in the gauge fixed form.
The symmetries in the superspace formulation are understood,
i.e. they are non-anomalous  or anomalous  depending whether the Berezinian
is equal or different from one, respectively.

We start with the non-anomalous  symmetries. In infinitesimal form the Berezinian being one gives the constraint
\bea\label{volumepreserving}
\Sigma(\lambda)={1\over3}\left(\pa_m \l^m - \pa_\a \l^\a\right)=0~.
\eea
These  non-anomalous transformations leave $H$ inside the gauge and are therefore  simply
residual gauge transformations. Their general (infinitesimal) form is
\begin{subequations} \label{4.14}
\bea
\l^m (\q) &=& a^m +2\ri \q \s^a \bar \e e_a{}^m -2 \q\e S^m
+\q^2 s^m\\
\l^\a (\q) &=& \e^\a +  \hf  \q^\a \pa_m a^m  - K^\a{}_\b \q^\b
-\q^2 \pa_m \Big[ \ri (\bar \e \tilde \s^a)^\a e_a {}^m +\e^\a S^m\Big]~,
\eea
\end{subequations}
where the  components obey the conditions
\bea
\bar a^m =a^m~, \qquad K_{\a\b} = K_{\b\a}~, \qquad \pa_m s^m =0~.
\eea
The parameters $a^m,\,K^\a{}_\b$ and $\e_\a$ correspond to general coordinate transformations,
local Lorentz transformations and local $Q$-supersymmetry transformations, respectively. They are all non-anomalous.
The identification of the parameters with the various symmetries proceeds by working out their
action on the components of $H^m$, using the infinitesimal form of \eqref{messH}, cf. \cite{OS80,BK}.
The gauge transformation generated by the complex transverse parameter $s^m$ acts
on the complex field $S^m$ as
\bea
\d_s S^m = s^m~,
\eea
while all other fields are invariant.
This identifies $S^m$ as the Hodge dual of a complex gauge three-form. Then
\bea
\mathcal{B} := \pa_m S^m
\eea
and its conjugate $\bar{\mathcal{B}}$ are the only independent gauge-invariant field strengths.
The fields $\big\{ e_a{}^m,\, \J^{m \a} ,\, \bar \J^{m}{}_\ad,\, A^m, \,\mathcal{B}, \,\bar{\mathcal {B}} \big\}$
constitute the multiplet of old minimal supergravity.

In  the parametrisation \eqref{4.14} for spacetime diffeomorphisms,
the component fields in \eqref{H-WZgauge2} are densities.
In order to deal with true tensor fields,
we have to switch to the following parametrisation \cite{OS80}
\bea
H^m (\q, \bar \q) &=& \frac{\ri}{2} e\Big( \q^2 S^m - \bar \q^2 \bar S^m \Big)
+ e \,\q \s^a \bar \q e_a{}^m
+ \ri \, e^{\frac 32} \Big(
\bar \q^2 \q^\a  \J^{m}{}_\a
-\q^2 \bar \q_\ad \bar \J^{m \ad } \Big) \non \\
&&
+ e^2 \, \q^2 \bar \q^2 \Big( A^m -\frac{\ri}{4 e} \Big[ S^n \pa_n (e \bar S^m)
- \bar S^n \pa_n (e S^m) \Big] \Big)~.
\label{H-WZgauge-rescaled2}
\eea
Some of the parameters in \eqref{4.14} should also be re-defined, in particular
\bea
s^m ~\to ~ e s^m~, \qquad \nabla_m s^m =0 \quad \Longleftrightarrow \quad
s^m = \ve^{mnrs } \nabla_n f_{rs}~.
\eea

We now discuss the $\lambda$-transformations with Berezinian different from one. Since they lead out of the
special gauge \eqref{H-WZgauge2}
they should be accompanied by a compensating gauge transformation. Alternatively their action could be calculated
using their unconstrained form using \eqref{lan} with $H$ having the special form. These transformations are all anomalous.
In particular they include $S$-supersymmetry, Weyl transformations and $R$-symmetry.
In addition there is the transformation
of the $S^m$ gauge field with transformation
\begin{equation}\label{Stran}
\delta_s S^m=s^m
\end{equation}
which in this scheme is anomalous if $\partial_m s^m\neq0$.


\section{The Wess-Zumino gauge and scheme}

Starting with the generating functional $\Gamma[H,\vf,\bar{\vf}]$, an extended class of ``ultralocal"
super-Weyl and $\lambda$-transformations allows us to reach the configuration
\bea
H^m_{\rm WZ} (\q, \bar \q) = \q \s^a \bar \q e_a{}^m + \ri \,\bar \q^2 \q^\a  \J^{m}{}_\a
-\ri \,\q^2 \bar \q_\ad \bar \J^{m \ad } +  \hf \q^2 \bar \q^2
\Big( A^m + \hf e_a{}^m \ve^{abcd} \o_{bcd} \Big)~,
\label{H-WZgauge}
\eea
and
\begin{equation}\label{vf}
\vf=\bar{\vf} =1~.
\end{equation}
This defines the WZ gauge, whose field content is that of conformal supergravity.
The transformations leading to this configuration are anomalous, so we cannot
have a ``gauge fixing" in the usual sense. We can, however, restrict $\Gamma$  to this configuration,
i.e. define an effective action in this gauge by
\begin{equation}\label{WZact}
\Gamma_{\rm WZ}[H_{\rm WZ}]\equiv\Gamma[H=H_{\rm WZ},\vf=1,\bar{\vf}=1]~.
\end{equation}
It is $\Gamma_{\rm WZ}$ which corresponds to the analysis carried out in \cite{KPST,Papa}.

Under infinitesimal $\lambda$ transformations
$\G_{\rm WZ}[H_{\rm WZ}]$ transforms as
\bea\label{WZaction-spurious}
\d_\l \G_{\rm WZ}[H|_{\rm WZ}]
=
-\int \rd^4 x \rd^2 \q  \,\S(\l) \langle T\rangle +{\rm c.c.}~,
\eea
where $\S(\l)$ is  defined in \eqref{B},
and the anomalous supertrace $\langle T \rangle$, eq. \eqref{s-Weyl-anomaly}, is evaluated at $\vf=1$ and $H=H_{\rm WZ}$.
Here we have to restrict the $\lambda$ transformations to those which preserve the WZ gauge for $H^m$.
They will be discussed at the end of this section.

Any further action of ultralocal  symmetries on $\Gamma_{\rm WZ}$ is known already for nonsupersymmetric theories
(like e.g. the shift of the Weyl anomaly to diffeomorphisms) so in the superconformal framework  the WZ gauge
represents an extremal situation. The exact action of the symmetries in the WZ gauge is completely fixed by the
algebra in superspace. Their restriction to the WZ gauge is  unique and it is valid independently of  any assumption
on the anomalies. A characteristic feature of this algebra is the local dependence of the structure
`constants' on the gauge fields, reflecting the compensating
gauge transformations needed to stay in the WZ gauge. The cohomological analysis of this algebra was done in
\cite{Papa} with the conclusion that in addition to  $S$-supersymmetry current, Weyl
transformations and $R$-symmetry gauge anomalies
there are anomalies also in the $Q$-supercurrent. There are exactly two cohomologically nontrivial solutions labelled
by the $a$ and $c$ coefficients.
In the Introduction we argued on general grounds that in such a situation one can ``uplift" the WZ gauge, i.e.
find a ``scheme" in superspace such that $\Gamma_{\rm WZ}$ is the gauge fixing
of $\Gamma$ suplemented by local counterterms which define the scheme.
We now construct this scheme explicitly.

Consider a WZ gauge configuration of the form \eqref{H-WZgauge} and perform a finite
$\lambda$-transformation defined by $f(y,\theta)$:
\begin{equation}\label{finite2}
y'^m=f^m(y,\theta)\equiv y^m-\theta^2 S^m(y)\, ,\quad \theta^{\prime\a}=\theta^{\a}\,,\quad
\bar{\theta}^{\prime\dot{\a}}=\bar{\theta}^{\dot{\a}}\,.
\end{equation}
Using the transformation rule
\begin{equation}\label{transf}
x'^m+iH'^m (x',\theta',\bar{\theta'})=f^m(x^m+i H_{\rm WZ})
\end{equation}
and its complex conjugate and the terminating expansion in $\theta^2$,  it is easy to show that
\begin{equation} \label{trick}
H'(x,\theta,\bar{\theta})=\tilde {H}(x,\theta,\bar\theta)
\end{equation}
where $\tilde{H}$ is the configuration defined in the minimal $Q$-supersymmetric scheme \eqref{H-WZgauge2}.
Moreover, the Berezinian of the transformation \eqref{transf} is
\begin{equation}\label{bere}
B=1+\theta^2 \partial_m S^m=1+\theta^2{\cal B}
\end{equation}
Then using the minimal $Q$-supersymmetric scheme \eqref{lan} we obtain
\begin{equation}
\tilde {\Gamma}[\tilde{H}]=\tilde{\Gamma}[H_{\rm WZ}]
-K[\tilde{H},1,1;\theta^2 {\cal B},\bar{\theta}^2 \bar{\cal B}]
\end{equation}
Combining \eqref{lan} and \eqref{H11} we obtain:
\begin{equation}\label{WZscheme}
\Gamma_{\rm WZ}[H_{\rm WZ}]=\Gamma[H,\vf,\bar{\vf}]+K[H,\vf,\bar{\vf};\vf^{-1},\bar{\vf}^{-1}]
+K[\tilde{H},1,1;\theta^2 {\cal B},\bar{\theta}^2 \bar{\cal B}]
\end{equation}
In the second $K$ term above we could replace $\tilde{H}$ with  the generic configuration $H$ by
using the $\lambda$-transformation group element $f_0(H)$.

The r.h.s. of \eqref{WZscheme}  depends only on $H_{\rm WZ}$, together with $\vf=\bar{\vf}=1$. This proves that the action
restricted to the WZ gauge corresponds to a ``scheme", i.e.  starting with the ``standard superspace scheme" represented
by $\Gamma[H,\vf,\bar{\vf}]$, we have added explicit local counterterms, i.e. the two $K$-functionals.
All the properties of the WZ gauge can now be read off from \eqref{WZscheme}. In particular the violation of
$Q$-supersymmetry is all in the second $K$-term  which has an explicit supersymmetry breaking as it contains
a superfield with explicit $\theta$ dependence. The anomaly in $Q$-supersymmetry does not have any
fundamental significance: it can be removed by simply adding a local counterterm
$-K[\tilde{H},1,1;\theta^2 {\cal B},\bar{\theta}^2 \bar{\cal B}] $.
We could be more specific about the shifting of the anomalies we used: comparing with the minimal $Q$-symmetric
scheme it is evident that Q-supersymmetry became anomalous when we gauged away the $S^m$ field,
even though the required transformation
\begin{equation}\label{Sano}
\delta S^m=s^m
\end{equation}
with $\partial_m s^m \neq 0$ is anomalous.
Therefore we simply shifted the anomaly from the transformation of $S^m$ to $Q$-supersymmetry.
In Appendix B we will discuss similar well known shifts of anomalies in non-supersymmetric theories.

Let us now discuss the residual symmetries which preserve the gauge \eqref{H-WZgauge2}. They are
parametrized by
\begin{subequations}\label{residual}
\bea
\l^m (\q) &=& a^m +2\ri \q \s^a \bar \e e_a{}^m
+2\q^2 \bar \e \bar \J^m ~, \qquad \bar a^m =a^m~, \\
\l^\a (\q) &=& \e^\a +\hf (\s + \ri \r ) \q^\a - K^\a{}_\b \q^\b  \non \\
&&\phantom{\e^\a}  +\q^2 \Big[ \eta^\a - \ri (\nabla_b \bar \e \tilde \s^b)^\a
+ (\bar \e \tilde{\s}_b)^\a
\Big( \frac{\ri }{2} \o_c{}^{cb} +\frac 14 \ve^{bcde} \o_{cde} \Big) \Big] ~.~~~
\eea
\end{subequations}
Here the component parameters correspond to
spacetime reparametrisations $(a^m$), local Lorentz ($K_{\ab}=K_{\b\a}$),
$Q$-supersymmetry ($\e^\a$), $S$-supersymmetry ($\eta^\a$),
local scale ($\s$) and $R$-symmetry ($\r$) transformations, i.e. the parameters of the
supercornformal group. The transformations under which $\Gamma_{\rm WZ}$ is non-anomalous,
i.e. which satisfy $\Sigma(\l)=0$, can be parametrized as
\bea
\l^m (\q) = a^m  ~, \qquad
\l^\a (\q) = \hf  \q^\a\pa_m a^m   + K^\a{}_\b \q^\b   ~.
\label{diffLor}
\eea
These are diffeomorphisms and local Lorentz transformations.

With this parametrisation \eqref{diffLor} for spacetime diffeomorphisms,
the component fields in \eqref{H-WZgauge} are no longer vector fields with respect to the index
`$m$', instead  they are vector densities. In order to work with true vector fields,
we have to switch to the following parametrisation (compare with \cite{OS80})
\bea
H^m (\q, \bar \q) &=& e \,\q \s^a \bar \q e_a{}^m + \ri \, e^{\frac 32} \Big(
\bar \q^2 \q^\a  \J^{m}{}_\a
-
\q^2 \bar \q_\ad \bar \J^{m \ad } \Big) \non \\
&&\qquad + \hf  e^2\, \q^2 \bar \q^2
\Big( A^m + \hf e_a{}^m \ve^{abcd} \o_{bcd} \Big)
~,
\label{H-WZgauge-rescaled}
\eea
with $e = \det (e_m{}^a)$.
The gauge freedom is then described by parameters
\begin{subequations}\label{residual2}
\bea
\l^m (\q) &=& a^m +2\ri  e^\hf\q \s^a \bar \e e_a{}^m
+2e \,\q^2 \bar \e \bar \J^m ~,
 \\
\l^\a (\q) &=& e^{-\hf} \e^\a
- \hf \Big(  3(\s+\ri\,\rho) -\pa_m a^m
+ \ri (\e\s_b \bar \J^n - \J^n \s_b \bar \e) e_n{}^b \Big) \q^\a
- K^\a{}_\b \q^\b  \non \\
&&  + e^\hf \q^2 \Big[ \eta^\a - \ri (\nabla_b \bar \e \tilde \s^b)^\a
+ (\bar \e \tilde{\s}^b)^\a
\Big(-\frac{\ri}{2}  e_b \ln e
+
 \frac{\ri }{2} \o^c{}_{cb} +\frac 14 \ve_{bcde} \o^{cde} \Big) \Big] ~,~~~
\eea
\end{subequations}
with $e_a = e_a{}^m \pa_m$. For $\S(\l)$ we obtain
\bea
\S(\l) &=& \s +\ri\,\r +\frac{\ri}{3}  (\e\s_a \bar \J^m - \J^m \s_a \bar \e) e_m{}^a
+\frac 23 e^{\hf} \Big( \q \eta + 2\ri \q \s^a \nabla_a \bar \e \Big)
 \non \\
&& \phantom{\s}
+\frac 23 e \q^2 \nabla_m (\bar \e \bar \J^m) ~.
\eea
which is independent of the parameters for general coordinate transformations $(a^m$) and
local Lorentz-transformations ($K^\a{}_\b$), the only remaining non-anomalous symmetries
of $\Gamma_{\rm WZ}$.

\section{Discussion}

While anomalies in supersymmetric theories obey the general constraints  of any relativistic QFT  following from
analyticity and unitarity, they have specific features caused mainly  by the proliferation of ``ultralocal null operators".
Related to that it  is tempting to study anomalies in gauges in which most of these operators are put to zero and
one concentrates  on the anomalies related to ``null divergences". We called these gauges WZ gauge generically.
In such a situation, when the operators are coupled to source gauge fields, the algebra of symmetry
transformations becomes field  dependent.

The cohomology problem for this algebra has nontrivial solutions where  locality, essential for the formulation
of the cohomology, is defined in terms of the still unfixed gauge fields. The dimension of the space of nontrivial solutions
of the cohomology problem is correctly obtained by this procedure but the characterization of the solutions in terms
of the necessarily anomalous symmetries is not always valid. The reason is that the  class of allowed local
counterterms is  much larger than the ones realized in the WZ gauges. Polynomials in momenta can be added to correlators
in the microscopic theory or, equivalently, local terms in the gauge field can be added to the effective action.
This is the case even when the operators are null, i.e. their correlators  do not have an imaginary part:  then the added
correlators are just polynomials or, equivalently, the added local terms contain gauge fields which are not really coupled in the
microscopic theory. This freedom is missed when one goes to the fixed gauge and therefore the possibility of  shifting the
anomaly from one symmetry to another is reduced.

A similar situation occurs already in non supersymmetric theories for trace anomalies, as we discuss in Appendix B.
The trace of the energy-momentum is an ``ultralocal null operator'' and we can go to a gauge where all its correlators are
put to zero. Then in this gauge an anomaly appears in the conservation of the energy-momentum tensor.
The dimension of the space of cohomologically nontrivial solutions (one type A and a number of space time dimension
dependent type B \cite{DSch})
is correctly reproduced, but it would be incorrect to conclude that the conservation of the
energy-momentum tensor is necessarily anomalous.
When one adds the ``Weyl mode" in the enlarged gauge field space, the anomaly can be shifted to the trace in a new scheme
and the conservation of the energy-momentum tensor
is reinstated: by adding pure polynomials in momenta for the correlators of the
trace the Ward Identities following from conservation are satisfied.

In  this paper we discussed in detail examples of  ${\cal N}=1$
supersymmetric theories in four dimensions. The superspace formulation is very convenient as
it provides  a large enough space of gauge directions or, equivalently, operators in the microscopic theory for which the local
counterterms could  appear. Then we could discuss systematically the restrictions which appear in a given gauge and
to what scheme they correspond in superspace. The relation between the superspace formulation and the WZ gauge is
an equivalence: to every solution of the cohomology problem in the WZ gauge there corresponds a scheme in superspace
such that the generating functional reproduces the exact dependence on the gauge fixed fields in the WZ gauge.
Then one can make changes in the scheme in superspace and thereby shift the anomaly.

In the flavour case there is a well-known apparent violation of global $Q$-supersymmetry in the WZ gauge.
In superspace, however, this turns out to be just another scheme where a non supersymmetric counterterm was
added to the action. Removing this term restores supersymmetry.

Microscopic superconformal models allow different partial gauge fixings. Having a scheme in superspace where the
super-Weyl mode is put to one is shown to be equivalent to fixing the gauge to the minimal one preserving non-anomalous
$Q$-supersymmetry. The field content in this gauge contains, in addition to the vierbein, gravitino and $R$-current gauge field,
an additional one form field $S_m$. This field content is enough to realize the anomalies for $S$-supersymmetry,
$R$-symmetry and Weyl invariance, leaving diffeomorphism invariance and $Q$-supersymmetry non-anomalous.

If one does an additional gauge fixing putting $S_m$ to zero, one reaches the WZ gauge. In this gauge the
$Q$-supersymmetry becomes anomalous. This can be understood as a result of shifting the anomaly in the $S^m$ shift
invariance to $Q$-supersymmetry.
Obviously the $Q$-supersymmetry anomaly can be removed by the opposite process.

One can continue to an even ``more physical'" gauge where  also the Weyl mode is fixed, in which case not only
$Q$-supersymmetry but also diffeomorphism invariance would look anomalous.

The common feature of the above examples is the apparent anomaly in $Q$-super\-symmetry.
The explicit realizations show exactly how the apparent anomalies in $Q$-supersymmetry were produced,
by simply choosing schemes where $Q$-symmetry violating local counterterms
were added to the effective action.

In terms of physical applications, an apparent anomaly in $Q$-symmetry would not change the consequences
of anomaly matching, provided the same gauge is used in the UV and IR. On the other hand, if one wants to make
supergravity fields dynamical, the matter (microscopic) theory should couple in an anomaly free way. Therefore,
while generically we will not be able to couple superconformal matter to dynamical superconformal gravity,
by choosing a scheme where $Q$-supersymmetry is non-anomalous we could couple to dynamical Einstein supergravity.

The analysis presented, in particular the nonexistence of anomalies in $Q$-supersymmetry relied on the existence
of a  superspace formulation. In principle this is not necessary: an analysis of  the correlators of the microscopic
theory taking into account all the possible ``ultralocal null operators" could replace it. In any case,
the existence of a nonremovable anomaly in $Q$-supersymmetry requires more solid arguments than just seing the anomaly in a
particular WZ-like gauge.

\noindent
{\bf Acknowledgements:}\\
SMK is grateful to the Albert Einstein Institute for hospitality
during an early stage of this project. His work is supported in part by the Australian
Research Council, project No. DP160103633. AS is grateful to the Albert Einstein Institute for its hospitality.
His work was supported by the Israel Science Foundation (grant number 1937/12).
Very useful discussions with O. Aharony, Z. Komargodski, S. Yankielowicz are gratefully acknowledged.
ST acknowledges helpful discussions with F. Ciceri, V. Prochazka and A. van Proeyen during the early stages of this project.

\appendix

\section{Generating the super-Weyl anomaly} \label{AppendixB}

The discussion of Section 4 used the existence of a local action (cf. \eqref{3.8}) whose Weyl variation
produces the anomaly. It was used to construct a counterterm which allowed us to go to
Wess-Zumino gauge without encountering an anomaly in the required symmetry transformations.
In this appendix we present an alternative derivation of this super-space effective action, which
was originally derived by intergrating the Weyl anomaly in superspace, using the procedure
of Wess and Zumino \cite{WZ-anomaly}.

As preparation we need to collect some facts about
our approach to $\cN=1$ conformal supergravity \cite{KTvN1,KTvN2} in superspace, which uses the
Grimm-Wess-Zumino geometry \cite{WZ,GWZ},
in conjunction with the super-Weyl transformations discovered
by Howe and Tucker  \cite{HT}.
They leave the algebra of supergravity covariant derivatives
\bea
\cD_A &=& (\cD_a , \cD_\a ,\cDB^\ad )
=E_A{}^M  \pa_M + \O_A{}^{\b \g} M_{\b \g}
+\O_A{}^{\bd \gd} {\bar M}_{\bd \gd}
\label{A.111}
\eea
invariant.
Here $M_{\b\g} =M_{\g\b}$  and ${\bar M}_{\bd \gd} ={\bar M}_{\gd \bd}$  are the Lorentz generators.
The algebra is given in eq. (5.3.53) in \cite{BK}, where other
details of the construction can also be found.

A super-Weyl transformation is
associated with a chiral parameter $\S$, $\bar \cD_\ad \S=0$ and its complex conjugate $\bar \S$. Its infinitesimal form is
\begin{subequations}
\label{superweyl}
\bea
\d_\S \cD_\a &=& ( \hf \S - {\bar \S} )  \cD_\a - \cD^\b \S M_{\a \b}  ~, \\
\d_\S \bar \cD_\ad & = & ( \hf {\bar \S} - \S  )
\bar \cD_\ad -   \bar \cD^\bd  {\bar \S}   {\bar M}_{\ad \bd} ~,\\
\d_\S \cD_{\a\ad} &=& -\hf( \S +\bar \S) \cD_{\a\ad}
-\frac{\ri}{2} \bar \cD_\ad \bar \S \cD_\a - \frac{\ri}{2}  \cD_\a  \S \bar \cD_\ad \non \\
&& - \cD^\b{}_\ad \S M_{\a\b} - \cD_\a{}^\bd \bar \S \bar M_{\ad \bd}~,
\eea
\end{subequations}
provided the torsion tensors transform as follows:
\begin{subequations}
\label{B2}
\bea
\d_\S R &=&
 (\bar \S - 2\S) R -\frac{1}{4} \bar \cD^2 \bar \S
~, \\
\d_\S G_{\a\ad} &=& -\hf (\S +\bar \S) G_{\a\ad} +\ri \,\cD_{\a\ad} (\bar \S- \S) ~,
\label{s-WeylG}\\
\d_\S W_{\a\b\g} &=&-\frac{3}{2} \S W_{\a\b\g}~.
\eea
\end{subequations}
Finally, $\delta_\Sigma E=(\Sigma+\bar\Sigma)E$ and $\delta_\Sigma{\cal E}=3\,\Sigma\,{\cal E}$.

Consider now the following functional
\bea\label{defI}
I &=&  \int \rd^{4|4}z \,E \, \bigg\{\ln (\O \bar \O )
(G^aG_a +2R \bar R)
- \frac 14 \Big[ R \cD^2 \ln \O + \bar R \bar \cD^2 \ln \bar \O \Big] \non \\
&&\qquad
+\hf G^{\a\ad} \bar \cD_\ad \ln \bar \O \cD_\a \ln \O
+  \frac{1}{16}
\cD^2 \ln \O \,  \bar \cD^2 \ln \bar \O
+R \bar R\bigg\}~.
\eea
Using the super-Weyl transformation of $\O$,  eq. \eqref{3.222},
the super-Weyl variation of $I$  is
\bea
\d_\S I = -  \int \rd^{4|4}z \,E \, (\S+\bar \S)(G^aG_a +2R \bar R)~.
\label{SWvariation}
\eea
To  prove this, we use results obtained in  \cite{ButterK2013,Butter:2013lta}.
First, if we convert the first line of \eqref{defI} into an integral over chiral superspace plus its complex conjugate,
we notice the combination $\X\,\ln\Omega$ where
\bea
\X:= -\frac{1}{4} (\bar \cD^2 -4R ) \Big\{ G^aG_a + 2R \bar R
- \frac{1}{4} \cD^2 R \Big\}
\eea
whose super-Weyl variation can be shown to be
\bea
&& \d_\S \X = -3\, \S\, \X + \D \bar \S~.
\eea
Here $\D$ denotes the following higher-derivative operator
\bea
\D\bar \F := -\frac{1}{64} (\bar \cD^2 -4R ) \Big\{ \cD^2 \bar \cD^2 \bar \F
+ 8 \cD^\a (G_{\a\ad}\bar \cD^\ad \bar \F)\Big\}~, \qquad
\bar \cD_\ad \D\bar \F =0~.
\label{B.7}
\eea
Next we use the relation 
\begin{equation}
\delta_\Sigma\Big\lbrace \cD^2\bar\cD^2\bar\Phi+8\,\cD^\a\big(G_{\a\dot\a}\bar\cD^{\dot\a}\bar\Phi\big)\Big\rbrace
=-(\Sigma+\bar\Sigma)\Big\lbrace\cD^2\bar\cD^2\bar\Phi+8\,\cD^\a\big(G_{\a\dot\a}\bar\cD^{\dot\a}\bar\Phi\big)\Big\rbrace
+\bar\cD_{\dot\a}\big(\dots\big)
\end{equation}
which is valid if $\delta_\Sigma\bar\Phi=0$. Finally we need
\begin{equation}
\delta_\S\int  \rd^{4|4}z \,E \, R\,\bar R=-{1\over4}\int  \rd^{4|4}z \,E \,(R\,\cD^2\Sigma+\bar R\,\bar\cD^2\bar\Sigma)
\end{equation}
The above observations are sufficient to prove \eqref{SWvariation}.

We can rewrite $I$ in a different form with the help of the identity
\bea
\bar \cD^2 \ln \bar \O = \frac{ (\bar \cD^2 -4R ) \bar \O}{\bar \O} + 4R
- (\bar \cD \ln \bar \O)^2
\eea
and its conjugate. Then $I$ becomes
\bea
I &=&  \int \rd^{4|4}z \,E \, \bigg\{\ln (\O \bar \O )
(G^aG_a +2R \bar R)
+\hf  G^{\a\ad} \bar \cD_\ad \ln \bar \O \cD_\a \ln \O
+  \frac{1}{16}
(\cD \ln \O)^2  ( \bar \cD \ln \bar \O)^2
\bigg\}\non \\
&&\qquad -\frac{1}{16} \int \rd^{4|4}z \,E \, \bigg\{
\frac{ ( \cD^2 -4\bar R )  \O}{\O}  ( \bar \cD \ln {\bar \O} )^2
+ \frac{ (\bar \cD^2 -4R ) \bar \O}{\bar \O} (\cD \ln \O)^2 \bigg\} \non \\
&& \qquad + \frac{1}{16} \int \rd^{4|4}z \,E \,
\frac{ ( \cD^2 -4\bar R )\O }{ \O} \frac{ (\bar \cD^2 -4R ) \bar \O}
{ \bar \O} ~.
\label{3.6}
\eea
The last line in above expression is Weyl-invariant and can be dropped. The resulting expression will be called $\tilde I$.
It is then clear that the combination
\begin{equation}
\G = -2(c-a)\int d^4 x\,d^2\theta\,{\cal E}\,\ln\Omega\,W^{\a\b\g}W_{\a\b\g}+{\rm c.c}+2\,a\,\tilde I
\label{A.12}
\end{equation}
solves \eqref{2.4}. Using the relations
\bea
-\frac{\ri}{4}  \int \rd^{4|4}z \,E \,
\cD^{\a\ad} \ln \O\,   \bar \cD_\ad \ln \bar \O\, \cD_\a \ln \O
= -\frac{1}{16}  \int \rd^{4|4}z \,E \,
(\cD \ln \O)^2 \bar \cD^2 \ln \bar \O
\eea
and its conjugate and
\bea
-\frac{1}{16} \int \rd^{4|4}z \,E \, \bigg\{
\frac{ ( \cD^2 -4\bar R )  \O}{\O}  ( \bar \cD \ln {\bar \O} )^2
&+& \frac{ (\bar \cD^2 -4R ) \bar \O}{\bar \O} (\cD \ln \O)^2 \bigg\}  \non \\
= \frac 14 \int \rd^{4|4}z \,E \, \bigg\{ R (\cD \ln \O)^2 + \bar R (\bar \cD \ln \bar \O)^2
&+&\ri
\cD^{\a\ad}\Big( \ln  \bar  \O - \ln \O  \Big) \bar \cD_\ad \ln \bar \O \cD_\a \ln \O \non \\
& - &\hf
(\cD \ln \O )^2 (  \bar \cD \ln \bar \O )^2 \bigg\}~.
\eea
one shows that \eqref{A.12} coincides with the functional $-K [H, \vf, \bar \vf; \O , \bar \O]$, see
\eqref{WZ}.

We remark that if we interpret $\Omega=\re^{-\Phi}$ as the dilaton superfield, as was done in \cite{SchT}, we can complete
the dilaton effective action obtained from 
$-K [H, \vf, \bar \vf; \O , \bar \O]$ by adding a Weyl invariant kinetic term
\bea
\int \rd^{4|4}z \, E\,\bar \O \O \,F
\Big( \frac{ (\bar \cD^2 -4R ) \bar \O}{\O^2}\, , \frac{ ( \cD^2 -4 \bar R ) \O}{\bar \O^2}
\Big) ~,
\label{DilatonEE}
\eea
where $F(\z, \bar \z)$ is a real function of one complex variable.
Ref. \cite{SchT} made the simplest choice $F(\z, \bar \z)=1$.

Given the above results, we can construct a non-local action which contains only
the supergravity fields $H^m$ and $\vf$. One possibility is to choose $\Omega$ such that
it satisfies the super-Weyl invariant massless equation
\begin{equation}\label{3.7}
(\cD^2-4\bar R)\Omega=0
\end{equation}
in which case the last two lines of \eqref{3.6} vanish.
The resulting effective action was constructed in \cite{BK-PLB88}.
More precisely, the chiral scalar $\O$ was chosen to coincide with
the unique solution to \eqref{3.7},  which was  proposed  in \cite{GNSZ}
as a non-local functional of the supergravity multiplet,
$\O = \O[ H, \vf , \bar \vf]$, and is given by 
\bea
\O = 1+ \frac{1}{4 \Box_+} ( \bar \cD^2 -4R) \bar R,
\label{boldomega}
\eea
where $\Box_+$ denotes the chiral d'Alembertian defined by
$\Box_+ \f = \frac{1}{16} ( \bar \cD^2 -4R) ( \cD^2 -4 \bar R) \f$,
for any covariantly chiral scalar $\f$.\footnote{This solution is a supersymmetric extension of
of the composite scalar field $\o= 1 + \frac{1}{6} (\Box - \frac{1}{6} \cR)^{-1} \cR$,
with $\cR$ the scalar curvature, proposed by Fradkin and Vilkovisky \cite{FV}.
The scalar field $\o$ was used by Fradkin and Tseytlin \cite{FT84} to integrate the ordinary
Weyl anomalies \cite{DDI,Duff,DSch}.}
With this choice of $\O$ the anomalous action \eqref{A.12} is 
the non-local effective action of \cite{BK-PLB88}
which we denote $\G^{\rm (I)}_{\rm anom}[H, \vf, \bar \vf]$.

There exist a different non-local effective action constructed in \cite{ButterK2013}, 
$\G^{\rm (II)}_{\rm anom}[H, \vf, \bar \vf]$, which makes use of the Green function
of the superconformal operator \eqref{B.7} and which
generates the super-Weyl anomaly. Both actions
$\G^{\rm (I)}_{\rm anom}[H, \vf, \bar \vf]$  and $\G^{\rm (II)}_{\rm anom}[H, \vf, \bar \vf]$
have the following fundamental properties: (i) they are manifestly locally supersymmetric and
(ii) they possess the same super-Weyl variation,
\bea
\d_\S \G^{\rm (I)}_{\rm anom}[H, \vf, \bar \vf] = \d_\S \G^{\rm (II)}_{\rm anom}[H, \vf, \bar \vf]= \d_\S \G[H, \vf, \bar \vf] ~,
\eea
with $\d_\S \G  [H, \vf , \bar \vf ]$ as in  eq. \eqref{2.4}.

It should be mentioned that the above choice for $\O$, eq. \eqref{boldomega},
is not unique. A slightly different chiral superfield was also used in \cite{BK-PLB88}
to integrate the super-Weyl anomaly. Interesting options emerge when we
consider a solution to the equations of motion of the dilaton effective action, 
see \cite{SchT} a discussion.

The important question of the analytic properties of these effective actions, i.e. whether they correctly reproduce the
correlation functions of the supercurrent multiplet, is not addressed here.
However, the `true' effective action differs from either one by at most a (non-local) super-Weyl invariant functional of $H^m$.


\section{Conformal  anomalies in the ``physical" gauge and scheme}

Conformal anomalies, in addition to being components of the superconformal anomalies,
present in a simplified setup the issues we faced in the main text. In a conformal theory, imposing the
equations of motion, the energy-momentum tensor
$T_{mn}$ is conserved while its trace  $T_{m}^{m}$ vanishes.
In special  situations, the free massless scalar in $d=2$ or Maxwell theory in $d=4$ , the  vanishing of the trace does not even require the equations of motion.
In every case the trace is an ultralocal null operator and therefore the system can be studied in
``physical" (analogue of Wess-Zumino) gauges using the ultralocality of gauge transformations.

In the general situation we couple the energy-momentum tensor to a metric $g$ on which the symmetries act:
diffeomorphisms related to conservation of the energy-momentum tensor and Weyl transformations related to its
tracelessness. The vanishing trace of the energy-momentum tensor means that not all components of the metric are
coupled and in a ``physical" gauge we could restrict the metric to a special class of  metrics
$\gg$ which obey $\det{\gg}=1$.

We start by studying the cohomology problem in this special gauge. The only symmetries left in this gauge are
spacetime diffeomorphisms parametrized by the infinitesimal parameters $\zeta^{m}(x)$.
The transformation of $\gg$ under diffeo  is given by
 \begin{equation}\label{diffeo}
\delta_{\zeta}\gg_{mn}=\hat{\nabla}_m \hat\zeta_{n}+\hat{\nabla}_{n}\hat\zeta_{m}
-{2 \over d}\hat{\nabla}\cdot\hat\zeta\, \gg_{mn}
\end{equation}
where covariant derivatives and raising and lowering indices are performed using the metric $\gg$, i.e.
$\hat\zeta_m=\hat g_{mn}\zeta^n$,  and the last
term is added in order to remain in the gauge after the transformation.

The cohomological problem is defined by asking for variations of  functionals $\hat{\Gamma}[\gg]$ such that
\begin{equation}\label{anomaly}
\delta_{\zeta}\hat{\Gamma}=\int d^{d}x \,\zeta^{m}\mathcal{A}_{m}(\gg)
\end{equation}
where $\mathcal{A}_{m}$ is local and the Wess-Zumino condition is obeyed:
\begin{equation}\label{wz}
\delta_{\zeta_1} \int d^{d}x\, \zeta^{m}_2 \mathcal{A}_{m}-\delta_{\zeta_2} \int d^{d}x\, \zeta^{m}_1 \mathcal{A}_{m}
=\int d^{d}x\, {(\zeta_1*\zeta_2)}^{m}\mathcal{A}_{m}
\end{equation}
where
\begin{equation}\label{Lie}
{(\zeta_1 *\zeta_2)}^{m}\equiv \zeta^{n}_1\partial_{n}\zeta^{m}_2- \zeta^{n}_2\partial_{n}\zeta^{m}_1
\end{equation}
The nontrivial solutions $\mathcal{A}_m$ of  \eqref{anomaly} and \eqref{wz} should be such that they do not
correspond to the variation of a local $\hat{\Gamma}$.	

In any even dimension there are  cohomologically nontrivial solutions given by
\begin{equation}\label{wano}
\mathcal{A}_{m}(\gg)=\partial_{m}\mathcal{A}(g=\gg)\,.
\end{equation}
Here $\mathcal{A}$ are the standard Weyl anomalies, i.e.
\begin{equation}\label{difano}
\mathcal{A}(g)=a\,E_d+\sum c_i\,W^i
\end{equation}
where $E_d$ is the $d$-dimensional Euler characteristic and $W_i$ are the Weyl invariant type B anomalies \cite{DSch}.
This will be discussed further below.

The anomaly modified  conservation equation is obtained by taking a functional derivative of
\eqref{anomaly} with respect to $\zeta^{m}$ and using the chain rule for the l.h.s. We obtain
\begin{equation}\label{emtens}
\hat{\nabla}^{m}\hat{T}_{mn}={1 \over 2}\partial_{n}\mathcal{A}(\gg)\,,
\end{equation}
where $\hat{T}$ is the automatically traceless energy-momentum tensor defined as
\begin{equation}\label{deff}
\hat{T}_{mn}\equiv{\delta \hat{\Gamma}\over \delta \hat{g}^{mn}}
-{1 \over d}\gg_{mn}\, \hat{g}^{rs}{{\delta \hat{\Gamma}} \over \delta\hat{g}^{rs}}\,.\
\end{equation}
The special form of \eqref{emtens} (with the gradient on the r.h.s.) indicates that the anomaly is ``removable",
i.e. the conservation can be reinstated  in another scheme. Indeed,  defining
\begin{equation}\label{remove}
\tilde{T}_{mn}\equiv \hat{T}_{mn}-{1\over2}\hat{g}_{mn}\mathcal{A}
\end{equation}
the new energy-momentum tensor will be conserved, but of course will not be  traceless.

The special form  of \eqref{emtens} is also related to the fact that while  diffeomorphism invariance
becomes anomalous in this scheme, the special class of diffeomorphisms
with unit determinant continue to be an anomaly free symmetry.
Consider for a generic diffeomorphism with parameters $\zeta^m$  the corresponding transformation with parameters:
\begin{equation}\label{traceless}
\tilde{\zeta}^m\equiv \zeta^m-\hat{\nabla}^m {1\over\hat{\square}} \hat{\nabla}\cdot\zeta\,.
\end{equation}
They have vanishing divergence.
We  expect that the variation of the action vanishes for these special transformations:
\begin{equation}\label{van}
0=\int d^{d}x\, \tilde{\zeta}^n \hat{\nabla}^{m}\hat{T}_{mn}\,.
\end{equation}
Using \eqref{traceless} we see that this is indeed the case, after integration by parts, provided
the r.h.s. of \eqref{emtens} is a gradient.

Following our general procedure we would now like to enlarge the space of couplings such that we can shift the anomaly 
away from diffeomorphisms. We add the Weyl mode $\Sigma$ coupled to the trace null operator. This produces an unconstrained 
metric $g_{mn}$ related to  the physical gauge metric $\gg_{mn}$ by
\begin{equation}\label{genm}
g_{mn}=(\exp2\Sigma)\,\gg_{mn}
\end{equation}
and
\begin{equation}\label{deter}
\exp{(2\,d\,\Sigma)}=\det{(g_{mn})}\,.
\end{equation}
We want to relate $\hat{\Gamma}$ to a generating functional $\Gamma$ in the enlarged space,  which depends on $g_{mn}$,
preserves diffeomorphism invariance and has an anomaly in the Weyl transformation of the metric
$\delta_{\sigma}g_{mn}=2\sigma g_{mn}$.
This generating functional $\Gamma[g_{mn}]$ has therefore the properties
\begin{equation}\label{diffeo1}
\delta_{\zeta}\Gamma=0
\end{equation}
and
\begin{equation}\label{ano1}
\delta_{\sigma}\Gamma= \int d^d x \sqrt{g}\, \sigma\, \mathcal{A}\,.
\end{equation}
For a finite Weyl transformation one has
\begin{equation}\label{wz2}
\Gamma[g_{mn}\exp{(2\sigma)}]=\Gamma[g_{mn}] +K[g_{mn};\exp{(2\sigma)}]~,
\end{equation}
where e.g. in $d=4$ \cite{Riegert}
\begin{equation}
\begin{aligned}
K[g_{mn};\exp(2\s)]&=a\int d^4 x\sqrt{g}\Big\lbrace\s\,E_4-4\big(R^{mn}-\textstyle{1\over2}g^{mn}R\big)\nabla_m\s\,\nabla_n\s
-4(\nabla\s)^2\square\s+2(\nabla\s)^4\Big\rbrace\\
&\qquad\qquad\qquad+c\int\sqrt{g\,}d^4 x\,\s\,C^2~.
\end{aligned}
\end{equation}}
Then $\hat{\Gamma}$ becomes the restriction of $\Gamma$, i.e.
\begin{equation}\label{restr}
\hat{\Gamma}(\gg)=\Gamma[{g_{mn} \det{(g)}}^{-{1\over d}}]~.
\end{equation}
The anomaly in diffeomorphisms following from the above definition  can be easily calculated since the variation
of the argument has automatically the form \eqref{diffeo}:
\begin{equation}\label{ano3}
\delta_{\zeta}\hat{\Gamma}=\Gamma[\gg+\delta_{\zeta}\gg]-\Gamma[\gg]~.
\end{equation}	
The first two terms in the variation \eqref{diffeo} correspond to a diffeomorphism transformation of $\Gamma$ under which it is
invariant while the third term is a Weyl transformation where we can identify the $\sigma$-parameter as
$-{1 \over d} \nabla.\zeta$. Using  \eqref{ano1} we obtain therefore
\begin{equation}\label{ano4}
\delta_{\zeta}\hat{\Gamma}=-{1 \over d}\int d^d x\ \sqrt{\gg}\, \nabla\cdot\zeta \mathcal{A}(\gg)\,.
\end{equation}
The restriction of $\Gamma$ to configurations $\gg_{mn}\equiv g_{mn}  (\det{g})^{-{1 \over d}}$ reproduces
the cohomology of $\hat{\Gamma}$ calculated directly in the ``physical"gauge.
Then it should be possible  to find a scheme in which a generating functional of $g_{mn}$ depends automatically only on
$\gg_{mn}$ i.e. . This should be a scheme where the $\sigma$ direction is not anomalous, thus
making the restriction to $\gg$ anomaly free. Such a scheme can be obtained using \eqref{wz2}
\begin{equation}\label{wz5}
\Gamma[g_{mn} {\det{(g)}}^{-{1\over d}}]=\Gamma[g_{mn}]+K[g_{mn};\det{(g)}^{-{1\over d}}]\,.
\end{equation}
 Defining
\begin{equation}\label{tilde1}
\tilde{\Gamma}\equiv\Gamma[g_{mn}]+K\big[g_{mn};\det{(g)}^{-{1\over d}}\big]
\end{equation}
we have a new scheme since $K$ is a local functional. Moreover $\tilde{\Gamma}$ is independent of the Weyl mode,
i.e. multiplying the metric becomes a genuine non-anomalous invariance. Therefore $\tilde{\Gamma}$
depends automatically only on $\gg$, which is a legal representative of the gauge orbit.
The breaking of diffeomorphism invariance is manifest by the appearance of the {\it scalar density} in the Wess-Zumino
functional.

The above treatment exemplifies the general pattern we discuss. One can start solving the cohomology problem in a ``physical"
gauge. Then one adds a ``spurious" direction which is coupled to an on-shell null operator such that the ``physical"
cohomology is a particular restriction of the ``spurious" direction. Using then the possibility of adding local counterterms,
i.e. changing the scheme, one arrives at a new generating functional for which the ``spurious" direction
is non-anomalous and the ``physical" gauge expression becomes equivalent to the generating functional in the new scheme.
The addition of the mode coupled to the null operator allowed for a systematic treatment of the local counterterms which are
needed to shift the diffeomorphisms anomaly seen in  the physical gauge. All the local counterterms in this case were
for amplitudes with vanishing imaginary parts, i.e.  pure polynomials in momentum space, in exact analogy to the situation 
discussed for supersymmetric models.


\begin{footnotesize}

\end{footnotesize}


\begin{thebibliography}{66}

\bibitem{PSS}
  O.~Piguet, K.~Sibold and M.~Schweda,
  ``General solution of the supersymmetry consistency conditions,''
  Nucl.\ Phys.\ B {\bf 174}  (1980) 183.

\bibitem{PS}
  O.~Piguet and K.~Sibold,
  ``The anomaly in the Slavnov identity for $N=1$ supersymmetric {Yang-Mills} theories,''
  Nucl.\ Phys.\ B {\bf 247}  (1984) 484.

\bibitem{BPT}
  L.~Bonora, P.~Pasti and M.~Tonin,
``Cohomologies and anomalies in supersymmetric theories,''
  Nucl.\ Phys.\ B {\bf 252}   (1985) 458.

\bibitem{GG}
  G.~A.~Girardi and R.~Grimm,
  ``Algebraic description of chiral anomalies and superspace geometry,''
  Nucl.\ Phys.\ B {\bf 912}  (2016) 224.

\bibitem{WB} J.~Wess and J.~Bagger,
{\it Supersymmetry and Supergravity},
Princeton Univ. Press, 1992.

\bibitem{GGRS}
S.~J.~Gates Jr., M.~T.~Grisaru, M.~Ro\v{c}ek and W.~Siegel,
{\it Superspace, or One Thousand and One Lessons in Supersymmetry},
Benjamin/Cummings (Reading, MA),  1983, hep-th/0108200.

\bibitem{BK} I.~L.~Buchbinder and S.~M.~Kuzenko,
{\it Ideas and Methods of Supersymmetry and
Supergravity or a Walk Through Superspace}, IOP, Bristol, 1995
(Revised Edition: 1998).

\bibitem{Nair}
H.~Itoyama, V.~P.~Nair and H.~c.~Ren,
``Supersymmetry anomalies and some aspects of renormalization,''
Nucl.\ Phys.\ B {\bf 262} (1985) 317.

\bibitem{Guadagnini}
E.~Guadagnini and M.~Mintchev,
``Chiral anomalies and supersymmetry,''
Nucl.\ Phys.\ B {\bf 269} (1986) 543.

\bibitem{Closset}
C.~Closset, L.~Di Pietro and H.~Kim,
``'t Hooft anomalies and the holomorphy of supersymmetric partition functions,''
arXiv:1905.05722 [hep-th].

\bibitem{KTvN1}
  M.~Kaku, P.~K.~Townsend and P.~van Nieuwenhuizen,
  ``Gauge theory of the conformal and superconformal group,''
  Phys.\ Lett.\  {\bf 69B} (1977) 304.

\bibitem{KTvN2}
  M.~Kaku, P.~K.~Townsend and P.~van Nieuwenhuizen,
  ``Properties of conformal supergravity,''
  Phys.\ Rev.\ D {\bf 17}(1978) 3179.

\bibitem{Howe}
P.~S.~Howe,
``A superspace approach to extended conformal supergravity,''
  Phys.\ Lett.\ B {\bf 100} (1981) 389;
``Supergravity in superspace,''  Nucl.\ Phys.\  B {\bf 199} (1982) 309.

\bibitem{Butter}
  D.~Butter, ``N=1 conformal superspace in four dimensions,''
  Annals Phys.\  {\bf 325} (2010) 1026 [arXiv:0906.4399 [hep-th]].

\bibitem{FT-conformal}
E.~S.~Fradkin and A.~A.~Tseytlin,  ``Conformal supergravity,''
Phys.\ Rept.\  {\bf 119}  (1985) 233.

\bibitem{GWZ}
  R.~Grimm, J.~Wess and B.~Zumino,
    ``Consistency checks on the superspace formulation of supergravity,''
  Phys.\ Lett.\ B {\bf 73} (1978) 415;
  ``A complete solution of the Bianchi identities in superspace,''
  Nucl.\ Phys.\ B {\bf 152} (1979) 255.

\bibitem{WZ}
J.~Wess and B.~Zumino,
``Superfield Lagrangian for supergravity,''
Phys.\ Lett.\  B {\bf 74}(1978) 51.

\bibitem{old1}
K.~S.~Stelle and P.~C.~West,
``Minimal auxiliary fields for supergravity,''
Phys.\ Lett.\  B {\bf 74} (1978)  330.

\bibitem{old2}
S.~Ferrara and P.~van Nieuwenhuizen,
``The auxiliary fields of supergravity,''
Phys.\ Lett.\  B {\bf 74} (1978) 333.


\bibitem{HT}
P.~S.~Howe and R.~W.~Tucker,
``Scale invariance in superspace,''
Phys.\ Lett.\ B {\bf 80}(1978) 138.

\bibitem{Siegel78}
W.~Siegel,
``Solution to constraints in Wess-Zumino supergravity formalism,''
Nucl.\ Phys.\  B {\bf 142} (1978) 301.

\bibitem{OS}
  V.~Ogievetsky and E.~Sokatchev,
  ``Structure of supergravity group,''
  Phys.\ Lett.\ B {\bf 79} (1978) 222.

\bibitem{Siegel:SC}
  W.~Siegel,
  ``Superconformal invariance of superspace with nonminimal auxiliary fields,''
  Phys.\ Lett.\ B {\bf 80} (1979) 224 .

\bibitem{GrisaruSiegel}
  M.~T.~Grisaru and W.~Siegel,
 ``Supergraphity (I). Background field formalism,''
  Nucl.\ Phys.\ B {\bf 187} (1981) 149;
 ``Supergraphity (II). Manifestly covariant rules and higher loop finiteness,''
Nucl.\ Phys.\ B {\bf 201} (1982)  292.


\bibitem{FZ75}
  S.~Ferrara and B.~Zumino,
``Transformation properties of the supercurrent,''
  Nucl.\ Phys.\  B {\bf 87} (1975) 207.

\bibitem{BK86}
  I.~L.~Buchbinder and S.~M.~Kuzenko,
  ``Matter superfields in external supergravity: Green functions, effective action and superconformal anomalies,''
  Nucl.\ Phys.\ B {\bf 274} (1986)  653.

\bibitem{McA}
I.~N.~McArthur,
``Super b(4) coefficients,''
  Phys.\ Lett.\ B {\bf 128}, 194 (1983);
``Super b(4) coefficients in supergravity,''
Class.\ Quant.\ Grav.\  {\bf 1}  (1984) 245.

\bibitem{SchT}
  A.~Schwimmer and S.~Theisen,
 ``Spontaneous breaking of conformal invariance and trace anomaly matching,''
  Nucl.\ Phys.\ B {\bf 847} (2011) 590
  [arXiv:1011.0696 [hep-th]].

\bibitem{Deser}
S.~Deser,
``Scale invariance and gravitational coupling,''
Annals Phys.\  {\bf 59} (1970)  248.

\bibitem{Zumino} B. Zumino,
``Effective Lagrangians and broken symmetries,"
in {\it Lectures on Elementary Particles and Quantum Field Theory,
Vol. 2}, S. Deser, M. Grisaru and H. Pendleton (Eds.),
Cambridge, Mass. 1970, pp. 437-500.

\bibitem{SG}
W.~Siegel and S.~J.~Gates Jr.
 ``Superfield supergravity,''  Nucl.\ Phys.\  B {\bf 147}(1979) 77.

\bibitem{deWR}
B.~de Wit and M.~Ro\v{c}ek,
``Improved tensor multiplets,''
Phys.\ Lett.\ B {\bf 109}  (1982) 439.

\bibitem{FGKV}
 S.~Ferrara, L.~Girardello, T.~Kugo and A.~Van Proeyen,
``Relation between different auxiliary field formulations of N=1 supergravity
coupled to matter,''   Nucl.\ Phys.\  B {\bf 223}  (1983) 191.

\bibitem{K13}
S.~M.~Kuzenko,
``Super-Weyl anomalies in N=2 supergravity and (non)local effective actions,''
JHEP {\bf 1310}  (2013) 151
[arXiv:1307.7586 [hep-th]].

\bibitem{deWG}
B.~de Wit and M.~T.~Grisaru,
``Compensating fields and anomalies,''
in  {\it Quantum Field Theory and Quantum Statistics, Vol. 2},
I. A.  Batalin, C. J. Isham and G. A. Vilkovisky (Eds.) Adam Hilger, Bristol, 1987, pp. 411--432.

\bibitem{SW1}
M.~F.~Sohnius and P.~C.~West,
``An alternative minimal off-shell version of N=1 supergravity,''
Phys.\ Lett.\  B {\bf 105} (1981) 353.

\bibitem{SW2}
M.~Sohnius and P.~C.~West,
``The tensor calculus and matter coupling of the alternative minimal auxiliary field formulation of $N=1$
supergravity,''  Nucl.\ Phys.\ B {\bf 198} (1982)  493.

\bibitem{non-min}
P.~Breitenlohner,
``A geometric interpretation of local supersymmetry,''
Phys.\ Lett.\  B {\bf 67} (1977)  49;
``Some invariant Lagrangians for local supersymmetry,''
Nucl.\ Phys.\ {\bf B124} (1977) 500.

\bibitem{Siegel-tensor}
W.~Siegel,
``Gauge spinor superfield as a scalar multiplet,''
Phys.\ Lett.\ B {\bf 85} (1979) 333.

\bibitem{GGS}
S.~J.~Gates Jr., M.~T.~Grisaru and W.~Siegel,
``Auxiliary field anomalies,''
Nucl.\ Phys.\ B {\bf 203} (1982) 189.

\bibitem{OS80}
V.~I.~Ogievetsky and E.~S.~Sokatchev,
``The simplest group of Einstein supergravity,''
Sov.\ J.\ Nucl.\ Phys.\  {\bf 31} (1980) 140
[Yad.\ Fiz.\  {\bf 31} (1980) 264].

\bibitem{KPST}
G.~Katsianis, I.~Papadimitriou, K.~Skenderis and M.~Taylor,
``Anomalous supersymmetry,''
Phys.\ Rev.\ Lett.\  {\bf 122}, no. 23  (2019) 231602
[arXiv:1902.06715 [hep-th]].

\bibitem{Papa}
I.~Papadimitriou,
``Supersymmetry anomalies in $\mathcal{N}=1$ conformal supergravity,''
JHEP {\bf 1904} (2019) 040
[arXiv:1902.06717 [hep-th]].

\bibitem{DSch}
S.~Deser and A.~Schwimmer,
``Geometric classification of conformal anomalies in arbitrary dimensions,''
Phys.\ Lett.\ B {\bf 309} (1993) 279
[hep-th/9302047].

\bibitem{WZ-anomaly}
J.~Wess and B.~Zumino,
``Consequences of anomalous Ward identities,''
Phys.\ Lett.\  {\bf 37B} (1971) 95.

\bibitem{ButterK2013}
D.~Butter and S.~M.~Kuzenko,
``Nonlocal action for the super-Weyl anomalies: A new representation,''
JHEP {\bf 1309}  (2013) 067
[arXiv:1307.1290 [hep-th]].

\bibitem{Butter:2013lta}
D.~Butter, B.~de Wit, S.~M.~Kuzenko and I.~Lodato,
``New higher-derivative invariants in N=2 supergravity and the Gauss-Bonnet term,''
JHEP {\bf 1312} (2013) 062
[arXiv:1307.6546 [hep-th]].

\bibitem{BK-PLB88}
I.~L.~Buchbinder and S.~M.~Kuzenko,
``Nonlocal action for supertrace anomalies in superspace of $N=1$ supergravity,''
Phys.\ Lett.\ B {\bf 202} (1988) 233.

\bibitem{GNSZ}
M.~T.~Grisaru, N.~K.~Nielsen, W.~Siegel and D.~Zanon,
``Energy-momentum tensors, supercurrents, (super)traces
and quantum equivalence,''  Nucl.\ Phys.\ B {\bf 247} (1984) 157.

\bibitem{FV}
E.~S.~Fradkin and G.~A.~Vilkovisky,
``Conformal off mass shell extension and elimination of conformal anomalies in quantum gravity,''
Phys.\ Lett.\ B {\bf 73} (1978) 209.

\bibitem{FT84}
E.~S.~Fradkin and A.~A.~Tseytlin,
``Conformal anomaly in Weyl theory and anomaly free superconformal theories,''
Phys.\ Lett.\ B {\bf 134} (1984) 187.

\bibitem{DDI}
S.~Deser, M.~J.~Duff and C.~J.~Isham,
``Nonlocal conformal anomalies,''
Nucl.\ Phys.\ B {\bf 111} (1976) 45.

\bibitem{Duff}
M.~J.~Duff,
``Observations on conformal anomalies,''
Nucl.\ Phys.\ B {\bf 125} (1977) 334.

\bibitem{Riegert}
R.~J.~Riegert,
``A Nonlocal Action for the Trace Anomaly,''
Phys.\ Lett.\  {\bf 134B} (1984) 56.

\end{thebibliography}
\end{document}